%% file: main.tex
\documentclass[10pt,journal,compsoc]{IEEEtran}
\ifCLASSOPTIONcompsoc
  \usepackage[nocompress]{cite}
\else
  \usepackage{cite}
\fi
\ifCLASSINFOpdf
\else
\fi
\include{includes}

\include{macros}

\begin{document}
%
\title{Neural Network Guided Evolutionary Fuzzing for Finding Traffic Violations of Autonomous Vehicles}
%
%
%
%

\author{Ziyuan Zhong,
        Gail Kaiser,
        Baishakhi Ray
\IEEEcompsocitemizethanks{\IEEEcompsocthanksitem Z. Zhong, G. Kaiser and B. Ray are with the Department
of Computer Science, Columbia University, New York,
NY, 10025.
E-mail: ziyuan.zhong@columbia.edu, kaiser@cs.columbia.edu, rayb@cs.columbia.edu
}
}

\input{body/abstract}

\maketitle

\IEEEdisplaynontitleabstractindextext

%
\IEEEpeerreviewmaketitle

\input{body/1_intro.v2}

\input{body/2_background}

\input{body/3_API}

\input{body/4_fuzzing}

\input{body/5_implementation}

\input{body/6_experiments}

\input{body/7_results}
\input{body/8_related}
\input{body/9_discussion}
\input{body/10_conclusion}

\input{body/11_ack}

\ifCLASSOPTIONcaptionsoff
  \newpage
\fi



%

\bibliographystyle{IEEEtran}

\bibliography{bib_main, bib_fuzzing}

\clearpage
\input{body/appendix}

%








\end{document}

%% file: includes.tex
\usepackage{amsmath}

\usepackage[font=small,labelfont=bf,tableposition=top]{caption}
\usepackage{subcaption}
\PassOptionsToPackage{hyphens}{url}\usepackage{hyperref}
\usepackage[noabbrev]{cleveref}
\hypersetup{colorlinks,urlcolor=blue,citecolor=black}
\usepackage[T1]{fontenc}
\usepackage[export]{adjustbox}
\usepackage[flushleft]{threeparttable}
\usepackage[linesnumbered, ruled]{algorithm2e}
\usepackage[utf8]{inputenc}
\usepackage{algorithmic}

\usepackage{amsthm}

\usepackage{amsfonts}
\usepackage[export]{adjustbox}

\usepackage{array}
\usepackage{balance}
\usepackage{bm}
\usepackage{blkarray}
\usepackage{booktabs}
\usepackage{color}
\usepackage{xcolor,colortbl}
\usepackage{comment}

\usepackage{enumitem}
\usepackage{eqparbox}
\usepackage{fancybox}
\usepackage{fancyvrb}
\usepackage{framed}
\usepackage{graphicx}
\usepackage{ifthen}
\usepackage{listings}
\usepackage{listofitems}
\usepackage{makecell}
\usepackage{mdwmath}
\usepackage{mdwtab}
\usepackage{microtype}
\usepackage{multirow}
\usepackage{ragged2e}
\usepackage{stackengine}
\usepackage{stfloats}
\usepackage{url}
\usepackage{wrapfig}
\usepackage{xspace}
\usepackage{nicefrac}
\usepackage{tabularx}

\usepackage{cleveref}

\usepackage{titlesec}
\usepackage{flushend}
\titlespacing*{\subsection}{0pt}{1pt}{1pt}
\titlespacing*{\paragraph}{0pt}{0pt}{1pt}

\usepackage{moreverb}

\everymath{\displaystyle}

\newcommand{\rom}[1]{\uppercase\expandafter{\romannumeral #1\relax}}

\newcommand{\etal}{\hbox{\emph{et al.}}\xspace}
\newcommand{\eg}{\hbox{\emph{e.g.,}}\xspace}
\newcommand{\ie}{\hbox{\emph{i.e.,}}\xspace}

\newcommand{\wrt}{\hbox{\emph{w.r.t.}}\xspace}

\newcommand{\etc}{\hbox{\emph{etc.}}\xspace}
\newcommand{\aka}{\hbox{\emph{a.k.a.}}\xspace}

\newcommand{\norm}[1]{\left\lVert#1\right\rVert}

\definecolor{gray50}{gray}{.5}
\definecolor{gray40}{gray}{.6}
\definecolor{gray30}{gray}{.7}
\definecolor{gray20}{gray}{.8}
\definecolor{gray10}{gray}{.9}
\definecolor{gray05}{gray}{.95}

\newlength\Linewidth
\def\findlength{\setlength\Linewidth\linewidth
\addtolength\Linewidth{-4\fboxrule}
\addtolength\Linewidth{-3\fboxsep}
}
\newenvironment{examplebox}{\par\begingroup
   \setlength{\fboxsep}{5pt}\findlength
   \setbox0=\vbox\bgroup\noindent
   \hsize=0.95\linewidth
   \begin{minipage}{0.95\linewidth}\normalsize}
    {\end{minipage}\egroup
    \textcolor{gray20}{\fboxsep1.5pt\fbox
     {\fboxsep5pt\colorbox{gray05}{\normalcolor\box0}}}
    \endgroup\par\noindent
    \normalcolor\ignorespacesafterend}

\newcounter{RQCounter}
\newcounter{RQACounter}

\newcommand{\RQ}[2]{%
\refstepcounter{RQCounter} \label{#1}
 \noindent
 \textbf{RQ\arabic{RQCounter}.~#2}
}

\newcommand{\RS}[2]{%
\begin{framed}%
\vspace{-1mm}
\noindent
\textbf{Result {\ref{#1}}:~}{\emph {#2}}%
\vspace{-1mm}
\end{framed}
}

\definecolor{javared}{rgb}{0.6,0,0} 
\definecolor{javagreen}{rgb}{0.25,0.5,0.35} 
\definecolor{javapurple}{rgb}{0.5,0,0.35} 
\definecolor{javadocblue}{rgb}{0.25,0.35,0.75} 

\definecolor{green}{rgb}{0.1,0.1,0.1}

\usepackage{listings}

\colorlet{punct}{red!60!black}
\definecolor{background}{HTML}{EEEEEE}
\definecolor{delim}{RGB}{20,105,176}
\colorlet{numb}{magenta!60!black}

\lstdefinelanguage{json}{
    basicstyle=\scriptsize\ttfamily\bf,
    numbers=none,
    numberstyle=\scriptsize,
    stepnumber=1,
    numbersep=8pt,
    showstringspaces=true,
    breaklines=true,
    frame=tb,
    stringstyle=\color{javared},
    backgroundcolor=\color{background},
    literate=
     *{0}{{{\color{numb}0}}}{1}
      {1}{{{\color{numb}1}}}{1}
      {2}{{{\color{numb}2}}}{1}
      {3}{{{\color{numb}3}}}{1}
      {4}{{{\color{numb}4}}}{1}
      {5}{{{\color{numb}5}}}{1}
      {6}{{{\color{numb}6}}}{1}
      {7}{{{\color{numb}7}}}{1}
      {8}{{{\color{numb}8}}}{1}
      {9}{{{\color{numb}9}}}{1}
      {:}{{{\color{punct}{:}}}}{1}
      {,}{{{\color{punct}{,}}}}{1}
      {\{}{{{\color{delim}{\{}}}}{1}
      {\}}{{{\color{delim}{\}}}}}{1}
      {[}{{{\color{delim}{[}}}}{1}
      {]}{{{\color{delim}{]}}}}{1},
}

\usepackage{amsmath}
\usepackage[linesnumbered,ruled]{algorithm2e}

\newcommand{\squishlist}{
 \begin{list}{\tiny$\blacksquare$}
 { \setlength{\itemsep}{0pt}
   \setlength{\parsep}{1pt}
   \setlength{\topsep}{1pt}
   \setlength{\partopsep}{0pt}
   \setlength{\leftmargin}{1.5em}
   \setlength{\labelwidth}{1em}
   \setlength{\labelsep}{0.5em} } }

\newcommand{\squishlisttwo}{
 \begin{list}{\tiny$\bullet$}
 { \setlength{\itemsep}{0pt}
  \setlength{\parsep}{0pt}
  \setlength{\topsep}{0pt}
  \setlength{\partopsep}{0pt}
  \setlength{\leftmargin}{1em}
  \setlength{\labelwidth}{1.5em}
  \setlength{\labelsep}{0.5em} } }
  
\newcommand{\squishend}{
 \end{list} }
 

%% file: macros.tex
\usepackage[normalem]{ulem}
\usepackage{cancel}

\newcommand{\new}[1]{\textcolor{red}{{#1}}}
\renewcommand{\new}[1]{\textcolor{black}{{#1}}}
\renewcommand{\sout}[1]{}

\newcommand{\remove}[1]{\textcolor{lightgray}{{#1}}}
\newcommand{\newnew}[1]{\textcolor{red}{{#1}}}

\renewcommand{\remove}[1]{}
\renewcommand{\newnew}[1]{{#1}}

\newcommand{\bray}[1]{}
\newcommand{\ziyuan}[1]{}
\newcommand{\gail}[1]{}

\newcommand{\newedit}[1]{\textcolor{black}{#1}}
\newcommand{\edit}[1]{\textcolor{black}{#1}}

\newcommand{\Comment}[1]{}

\newcommand{\tool}{\textit{AutoFuzz}\xspace}

\newcommand{\carla}{\textsc{carla}\xspace}
\newcommand{\CARLA}{\textsc{carla}\xspace}

\newcommand{\apollo}{\textsc{Apollo6.0}\xspace}
\newcommand{\svl}{\textsc{SVL}\xspace}

\newcommand{\allx}{\hbox{$\mathcal{X}$}\xspace}
\newcommand{\ally}{\hbox{$\mathcal{Y}$}\xspace}
 \newcommand{\av}{\hbox{\textsc{av}}\xspace}

\newcommand{\bugs}{traffic violations\xspace}
\newcommand{\bug}{traffic violation\xspace}

\newcommand{\NSGA}{\textsc{nsga}{\smaller2}\xspace}

\newcommand{\RA}{\textsc{random}\xspace}
\newcommand{\GA}{\textsc{ga}\xspace}
\newcommand{\NSGASM}{\textsc{nsga}{\smaller2}\textsc{-sm}\xspace}

\newcommand{\NSGADT}{\textsc{nsga}{\smaller2}\textsc{-dt}\xspace}

\newcommand{\GAUN}{\textsc{ga-un}\xspace}

\newcommand{\NSGAUNSMA}{\textsc{nsga}{\smaller2}\textsc{-un-sm-a}\xspace}

\newcommand{\GAUNNN}{\textsc{ga-un-nn}\xspace}
\newcommand{\GAUNNNGRAD}{\textsc{ga-un-nn-grad}\xspace}

\newcommand{\RAUNNNGRAD}{\textsc{random-un-nn-grad}\xspace}

\newcommand{\AVFUZZER}{\textsc{AV-FUZZER}\xspace}

\newcommand{\fss}{{functional scenarios}\xspace}
\newcommand{\ls}{{logical scenario}\xspace}
\newcommand{\lss}{{logical scenarios}\xspace}
\newcommand{\specifics}{{specific scenario}\xspace}
\newcommand{\specificss}{{specific scenarios}\xspace}
\newcommand{\is}{{initial scene}\xspace}
\newcommand{\iss}{{initial scenes}\xspace}

\makeatother

%% file: body/abstract.tex
\IEEEtitleabstractindextext{%

\begin{abstract}

Self-driving cars and trucks, autonomous vehicles (\av{}s), should not be accepted by regulatory bodies and the public until they have much higher confidence in their safety and reliability --- which can most practically and convincingly be achieved by testing. But existing testing methods are inadequate for checking the end-to-end behaviors of \av{} controllers against complex, real-world corner cases involving interactions with multiple independent agents such as pedestrians and human-driven vehicles. While test-driving \av{}s on streets and highways fails to capture many rare events, existing simulation-based testing methods mainly focus on simple scenarios and do not scale well for complex driving situations that require sophisticated awareness of the surroundings. To address these limitations, we propose a new fuzz testing technique, called \tool, which can leverage widely-used \av{} simulators' API grammars\new{ to}\sout{.
To} generate semantically and temporally valid complex driving \new{scenarios}\sout{secnarios} (sequences of scenes). \new{To efficiently search for \bugs-inducing scenarios in a large search space, we propose a constrained neural network (NN) evolutionary search method to optimize \tool.}\sout{\tool is guided by a constrained Neural Network (NN) evolutionary search over the API grammar to generate scenarios seeking to find unique \bugs.} 
Evaluation of our prototype on one state-of-the-art learning-based controller, two rule-based controllers, and one industrial-grade controller \new{in five scenarios} shows that \tool efficiently finds hundreds of \bugs in high-fidelity simulation environments. \new{For each scenario, \tool can find on average 10-39\% more unique \bugs than the best-performing baseline method.} Further, fine-tuning the learning-based controller with the \bugs found by \tool successfully reduced the \bugs found in the new version of the \av{} controller software. 

\end{abstract}

}

%% file: body/1_intro.v2.tex
\section{Introduction}

The rapid growth of autonomous driving technologies has made self-driving cars around the corner. 
As of June 2021, there are 55 autonomous vehicle (\av) companies 
actively testing self-driving cars on public roads in California~\cite{selfdrivingcompanies}. However, the safety of these cars remains a significant concern, undermining wide deployment --- there were 43 reported collisions involving self-driving cars in 2020 alone that resulted in property damage, bodily injury, or death \cite{selfdrivingcollisions}. 
Before mass adoption of \av for our day-to-day transportation, it is thus imperative to conduct comprehensive testing to improve their safety and reliability.

However, real-world testing (\eg monitoring an \av{} on a regular road) is extremely expensive and may fail to test against realistic variations of corner cases. Simulation-based testing is a popular and practical alternative~\new{\cite{blackbox19, Gambi2019Generating, Abeysirigoonawardena2019Generating, testing_vision18}}\sout{\cite{blackbox19, Gambi2019Generating, Abeysirigoonawardena2019Generating, ding2020learning, testing_vision18, nsga2nn, interactiontest}}. In a simulated environment, the main \av{} software, known as the {\em ego car controller}, receives multi-dimen\-sional inputs from various sensors (\eg Cameras, LiDAR, Radar, \etc) and processes the sensors' information to drive the car.

\begin{table}[h]
    \vspace{5pt} 
    \centering
    \scriptsize
     \setlength{\tabcolsep}{3pt}
     \renewcommand{\arraystretch}{1}
    \caption{\textbf{\small{Dominant Scenarios Leading to Car Crashes as per National Highway Traffic Safety Administration (NHTSA) report~\cite{precrash}.
    }}}
    \label{tab:crash}
  \vspace{-5pt}
    \begin{tabular}{l|l|l|l}
    \toprule
        Crash & \# Per & Economic & Years\\
        Scenario & Year  & Cost &  Lost \\
    \toprule
    A leading vehicle stopped & 975k & \$15,388m & 240k \\
    \sout{Ego car}\new{Vehicle} lost control without taking any action 
        & 529k & \$15,796m & 478k \\
    Vehicle(s) Turning at Non-Signalized Junctions & 435k & \$7343m & 138k \\
    A leading vehicle decelerating & 428k & \$6390m & 100k \\
    \sout{Ego car}\new{Vehicle} drove off road without taking any action  & 334k & \$9005m & 270k\\
    Straight Crossing Paths at Non-Signalized Junctions & 264k & \$7290m & 174k \\
    \bottomrule
    \end{tabular}
    {\scriptsize \sout{The car controlled by the user (through physical controls or an \av software) is commonly called the `\emph{ego car}'.} `\emph{Without taking any action}' \sout{here }means the \new{vehicle}\sout{ego car} is going straight or negotiating a curve \sout{rather }than explicitly making turns / changing lanes / leaving a parking position.
    } 
    \vspace{-1mm} 
\end{table}

A good simulation-based testing framework should test the ego car controller by simulating \new{challenging }real-life situations\sout{ that may lead to \bugs} --- especially the ones that emulate real-world violations made by human drivers that lead to crashes, such as those shown in~\Cref{tab:crash}. These crash scenarios are rather involved, \eg a leading car suddenly stopped to avoid a pedestrian and got hit by \new{a following vehicle.}\sout{the ego car from behind.} 
However, simulating such involved crash scenarios is non-trivial, especially because the ego car can interact with its surroundings (\new{\eg} driving path\sout{, road condition}, weather, stationery, and moving agents, etc.) in an \sout{exponentially large}\new{exponential} number of ways. 
Yet, simulating {\em some} crash-inducing scenario\new{s}, even in this large space, is not so difficult---for example, one can simply place a stationary object on the ego car's path to simulate a crash. 
Further, many \bugs can be reported with slight variations of essentially the same situation (\eg changing \new{a}\sout{an} never seen object's color). 
\gail{color is not a good example, since there are well-known av crashes that happened because another vehicle was white - so changing another vehicle from, say, blue to white really can matter.}
\ziyuan{I added "never seen"}
Thus one of the requirements for a successful simulation-based testing framework is to simulate scenarios that can lead to many {\em diverse} violations.

For traditional software, fuzz testing (a.k.a. fuzzing)~\new{\cite{fuzzing-book, FuzzTesting}}\sout{\cite{fuzzing-book, miller1990empirical, FuzzTesting}} is a popular way to find diverse bugs by navigating large search spaces. 
At a high level, fuzzing mutates existing test cases to generate new tests with an objective to discover new bugs.
\gail{maybe add something about work on fuzzing that specifically addresses the diversity of those bugs? it is certainly possible to use fuzzing to find a bunch of bugs that are essentially the same, fuzzing does not automatically bring diversity}
\ziyuan{I feel that we might want to not introducing extra work regarding diversity since it will influence the flow here.}
However, incorporating fuzzing into simulation testing of \av is not straightforward, as the test inputs (\ie driving scenarios in our case) have many features and inter-dependencies, and random mutations of arbitrary features will lead to semantically incorrect scenarios.  
\gail{the same would be true of any other fuzzing of structured data, i presume nearly all modern fuzzers pay attention to valid vs. invalid structure/semantics.}
\ziyuan{I am not sure if this is an issue since these structure/semantics are application-specific and our contribution is in the \av application.}
Although the simulator will eventually reject such inputs, the computational effort \sout{spent }on generating and validating these invalid test cases will waste a large portion of the testing budget. Thus, each generated scene and sequence of scenes (a scenario consists of a sequence of scenes) should be {\em semantically correct} as well as triggering {\em diverse} \bugs.






\begin{figure}[t]
    \centering
    \includegraphics[width=0.48\textwidth]{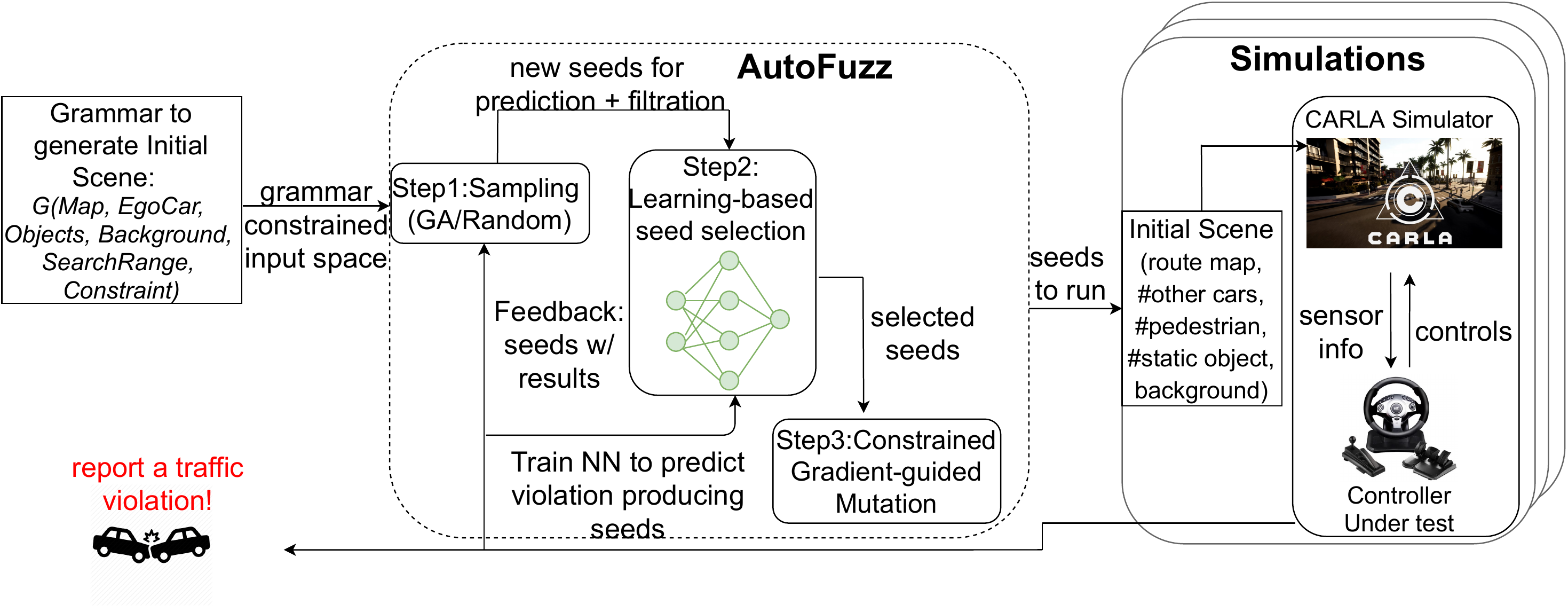}
    \vspace{-15pt}
    \caption{\tool High-level Overview}
    \label{fig:workflow}
\end{figure}

\smallskip
\noindent
\textbf{Our Approach.} We address these challenges by designing a gra\-m\-mar-guided learning-based fuzzer, called \tool\sout{, illustrated in} \new{(}\Cref{fig:workflow}\new{)}.
A self-driving car simulator takes some valid \edit{initial} scene configuration as input (consisting of: road map; starting position and destination of the ego car; initial locations, directions, and velocities of other cars and pedestrians; \etc) and starts the simulation with the \edit{initial} scene to generate a series of semantically valid consecutive scenes in the constrained driving environment. For \edit{initial scene} generation, \tool leverages the API grammar provided by the simulator and fuzzes the grammar-constrained input space, treating the simulator as black-box (\cref{sec:fuzzing}). 
\gail{the following sounds really incremental: the evolutionary fuzzing is 'typical' and we 'follow previous work on testing av systems'}
\ziyuan{I removed "typical". not sure if we should remove "following previous work..."}
In particular, \tool runs in \edit{an evolutionary} fuzzing setting where it is optimized to generate test input that the target simulator uses to initiate a scenario, running the ego car through corresponding time steps such that it may lead to a \bug. However, if we optimize the search to only find violation-producing inputs (\ie binary objective), it will be challenging to converge in a sparse space. Instead, following previous work on \new{\av{} }testing\sout{ \av{} systems}~\cite{interactiontest, Kuutti2020, nsga2nn, testing_vision18}, we formulate the fuzzing process as a smooth multi-objective search that guides the ego car\sout{ approach} to the point of interest.


To quantify the notion of \bug diversity, we define the concept of {\em unique violation}, where the configurations of two violation-producing input scenes should be apart by a user-defined threshold. 
\tool is optimized towards finding unique violations rather \edit{than every possible \bug}.
\gail{sounds like it finds every possible bug, just doesn't report, which wastes lots of simulation time}
\ziyuan{I removed the word "reporting"}
However, unique violation-producing inputs are sparse, and sparsity increases as the uniqueness threshold becomes more stringent. 
In such a sparse domain, the success of a fuzzer depends heavily on its initial seed selection and mutation strategy~\cite{she2020mtfuzz}, as successful mutants are often limited in a sparse high-dimensional space, and chances of finding them without any guidance are thin. \new{Besides, when a violation has been found, it is not trivial to automatically derive new violations with different parameters since a specific scenario leading to a violation can be very similar to a specific scenario leading to a safe outcome. One example is shown in \Cref{fig:variation_demo} where a small change of the leading vehicle's speed can lead to drastically different results.} To address \new{these}\sout{this}, we propose a novel seed selection and mutation strategy. Our key insight is, we can learn from the success/failure of the past mutants to produce \bugs and incorporate that knowledge in our fuzzing strategy. In particular, we devise a novel (i) learning-based seed selection and (ii) a gradient-guided mutation strategy that exploits knowledge learned from previous simulations. 
\gail{most work on mutation-based anything learns from the success of previous mutants in doing whatever they were supposed to do.  this is the basis of all genetic algorithms, i.e., evolutionary approaches to anything.}

\textit{Seed Selection.} \tool learns from previous test-runs' behavior in an incremental learning setting and leverages past knowledge to filter out new test cases (\aka seeds) that are unlikely to produce unique \bugs.  In particular, at each generation, we train a Neural Network (NN) classifier \new{\cite{nn_book98, GoodBengCour16, she2018neuzz, she2020mtfuzz}} on previous runs' results to predict if a new input will lead to a unique \bug. The confidence scores of the NN's prediction are then used to rank the candidate inputs from highest to lowest, with the top ones are selected.

\textit{Mutation Strategy.} The selected seeds are further mutated to increase their likelihood of causing unique \bugs. Here we leverage a projected gradient descent (PGD) \cite{DBLP:conf/iclr/MadryMSTV18} strategy from the ML-based adversarial attack domain. At a high level, a small {mutation} is added to \edit{every relatively lower confident input} from the seed selection step \edit{to increase the NN's confidence in it}, by iteratively back-propagating the NN's gradient. 
\gail{why are we even bothering with the lower confidence inputs?  above it says only the top ones are taken.}
\ziyuan{There are still relatively lower confident input among the most confident input and we mutate them to make them more confident.}
\ziyuan{I removed "to increase diversity" from "to increase the NN's confidence in it to increase diversity"}
However, naively applying gradient-guided mutation can generate invalid {inputs}. We resolve this problem by projecting each mutation back into a feasible region. \sout{Essentially, t}\new{T}he projection finds a feasible {mutation value} that obeys the grammar constraints and is also closest to the original {mutation value}. For this \tool applies a gradient-guided linear regression, where the grammar constraints are expressed as linear equations and the {corresponding fields of the mutation values} are variables. 

\new{Compared with previous works using evolutionary search based methods for \av testing \cite{avfuzzer, nsga2nn, testing_vision18}, our proposed seed selection and mutation strategy enable \tool to find more unique \bugs. Besides, unlike previous works which focus on one particular (mostly proprietary) system in a couple of fixed scenarios running in a particular simulator, we show the effectiveness of our proposed open source fuzzer \tool in the combination of multiple \av controllers, scenarios, and simulators.}
\smallskip
\noindent
\sout{This paper makes}\new{In summary, we make} the following contributions: 
\begin{itemize}[leftmargin=*,noitemsep,topsep=0pt]
\item We introduce \tool, a grammar-based fuzzing technique to test \av controllers, which leverages the simulator's API specification to generate semantically valid test scenarios.
\item \new{We propose a novel learning-based seed selection and mutation strategy to optimize \tool for finding more unique \bugs.}\sout{We optimize \tool to find unique \bugs using a novel learning-based seed selection and mutation strategy.}
\item \new{We evaluate our \tool prototype on four \av controllers \cite{chen2019lbc, ding2020learning, apollo} in two simulators \cite{carla, svl}. On average, \tool can find 10-39\% more unique \bugs per scenario than the best-performing baseline method.} \sout{We evaluate our \tool prototype for the widely used \carla simulator~\cite{carla}, on one end-to-end learning-based \av controller \cite{chen2019lbc} and two rule-based controllers \cite{chen2019lbc, ding2020learning}, and report hundreds of \bugs.}
\item We reduce \bugs by 75-100\% for the learning-based controller by fine-tuning it with the \bug-producing test cases. \sout{We show that \tool can also find \bugs using another simulator \svl~\cite{svl,svl-paper} with an industrial-grade controller \apollo~\cite{apollo}.}
\item We make \tool's source code and representative \bugs available at \url{https://github.com/autofuzz2020/AutoFuzz}\new{\cite{ziyuan_zhong_2022_6399383}}.
\end{itemize}

\smallskip
\noindent
\textbf{Contribution to SE Field.} 
\new{First, the proposed seed selection and mutation strategy can be potentially applied to other fuzzing areas where inputs take a long time to execute, and one needs to leverage time and effectiveness. Second, \tool is the first open source general framework on fuzz testing for \av{}s in high fidelity simulators. It allows a user to test a new system under a user-specified scenario in popular, open-source high-fidelity simulators. Besides, it allows a researcher to compare a new \av fuzzing method with existing methods easily. We believe the paper along with \tool can make the research in the field of \av testing more accessible and efficient to the community.}
\sout{This paper is core to the software testing field, particularly test generation, in our case for testing self-driving car controllers. We also show the potential of improving the controller software, thus contributing to the automated software repair literature. We hope this paper will overall improve the reliability of \av{}s.}


%% file: body/2_background.tex
\section{Background}
\label{sec:background}
\subsection{Definitions}
\label{sec:definition}

%



\noindent
First, we define a few terms \new{based on}\sout{, most of them taken from} \cite{scenescenariodef, scenariodefinitions}:


\noindent 
A \textbf{Scene} is a frame in the simulation that contains the detailed properties (\eg location, velocity, acceleration) of the ego-car, other moving objects, the surrounding stationary objects, and road conditions.  For example, the ego car is at map location (20, 20) with speed 5 m/s facing north on a rainy afternoon. 





\noindent
A \textbf{Scenario} is ``the temporal development between several scenes in a sequence of scenes'' \cite{scenescenariodef}. Two scenes could specify the same initial locations for the ego-car and other objects but different velocities, \etc resulting in different scenarios.  


\noindent 
A \textbf{Functional Scenario} is a natural language description of an abstract scenario, \eg the ego-car crosses an intersection. The examples in~\Cref{tab:crash} belong to this category. Since such an abstract functional scenario cannot be fuzzed directly, we design a corresponding logical scenario as a special implementation of the former. 
\noindent
A \textbf{Logical Scenario} is the parameterized space where search during the fuzzing will be bounded. For example,  the ego car that is crossing the intersection in the above example will start and end at locations $(x_s, y_s)$ and $(x_e, y_e)$, respectively, where $x_s, y_s \in [0, 20]$ and $x_e, y_e \in [20, 40]$. 



\noindent 
A \textbf{Specific Scenario} is a concrete instance \sout{to simulate sampling from}\new{in} the logical search space, \eg the ego car crossing the intersection will start at $(10, 10)$ and end at $(30, 30)$.  
A \specifics usually takes 30-50 seconds---if the simulation runs at 10Hz, this gives around 300-500 consecutive scenes.

\subsection{Testing Autonomous Vehicle  Controllers}
\label{sec:background_test}

There are three ways to test a controller: real-world, individual component, and simulation. 

\textbf{Real-world testing} involves running the controller on the road. 
\sout{For example, Waymo has tested its cars on public roads for 20 million miles from 2009 to 2018~\cite{waymoreport}, which is far less than the average yearly total driving distance in the U.S. (3 trillion miles per year) \cite{drivingmiles, drivernumber}.}
However, as per~\Cref{tab:crash}, many pre-crash functional scenarios may only occur in certain corner cases, \ie, variations in\sout{ road conditions,} background buildings, weather,\sout{ lighting,} the behaviors of other vehicles\sout{ and pedestrians}, \etc
It is extremely difficult to focus real-world testing towards such rare events. 


\textbf{Single component testing} primarily focuses on the perception component \sout{\cite{Eykholt2018robust, Zhao2019seeing, Zhou2018deepbillboard, Jia2020Fooling, Cao2019adversarial, Tian2018deeptest, Zhang2018deeproad}} 
or the planning component\sout{\cite{alessandro20, paolo21, luo21}}. The works for the perception component differ on the place perturbed: road sign
\new{\cite{Zhao2019seeing}}\sout{\cite{Eykholt2018robust, Zhao2019seeing}}, billboard\cite{Zhou2018deepbillboard}, LiDAR input\new{\cite{Jia2020Fooling}}\sout{\cite{Jia2020Fooling, Wang2021AdvSimGS}}, camera image\cite{Cao2019adversarial, Tian2018deeptest, Zhang2018deeproad}), LiDAR and camera image\cite{tumultisensor2021}, and the target they attack: perception\new{\cite{Zhao2019seeing, Zhou2018deepbillboard, Jia2020Fooling, Cao2019adversarial}}\sout{\cite{Eykholt2018robust, Zhao2019seeing, Zhou2018deepbillboard, Jia2020Fooling, Cao2019adversarial}}, motion planing\cite{Wong2020TestingTS}, lane following controller\cite{Tian2018deeptest, Zhang2018deeproad}. The works for the planning component differ on the characteristics of the scenarios to look for: avoidable collisions\cite{alessandro20}, patterns satisfaction\cite{paolo21}, and requirements violation\cite{luo21}.
However, this line of research tends to miss more involved interactions between different components~\cite{fusionattack}. 

\textbf{Simulator-based end-to-end testing} treats the ego-car controller as an end-to-end system and usually uses high-fidelity simulations to find failure cases. 
\new{Gambi et al. \cite{Gambi2019Generating} create simulations that reproduce \specificss according to the \fss leading to real car crashes in police reports. However, their system does not support testing different variations of the constructed \specificss, which is important to test for corner case behavior.}
\sout{There are three main ways: (i) constructing a known-hard testing-specific scenario \cite{Gambi2019Generating}, (ii) adding noise to sensor inputs \cite{fuzztest20}, and (iii)} 
\new{Most other works study how to efficiently find challenging} \sout{searching known-hard} specific scenarios in a parameterized logical scenario space.
\sout{\cite{zhong2021detecting, pathgeneration, blackbox19, NEURIPS2018_653c579e, wheelerimportance2019, Abeysirigoonawardena2019Generating, ding2020learning, Kuutti2020, chenbaiming2020, korenadaptive2018, multimodaltest, testing_vision18, interactiontest, nsga2nn, paracosm, avfuzzer, zhong2021survey}.} 
\sout{Gambi et al. \cite{Gambi2019Generating} create simulations that reproduce \specificss according to the \fss leading to real car crashes in police reports. However, their system does not support testing different variations of the constructed \specificss, which is important to test for corner case behaviors.
Han et al.\cite{fuzztest20} apply fuzz testing by randomly adding static boxes into the controller's sensor. Such tests cannot capture dynamic agents, \eg pedestrians.}\sout{Many works of the third category} \new{These works} usually model the \ls with only one or two agents having relatively simple behavior. However, many real-world crashes involve multiple dynamic agents with involved interaction (\eg a leading car brakes when the ego car gets close within a certain distance). 
Further, these works usually focus only on collisions \edit{rather than other \sout{types of }\bugs like going off-road}.
Furthermore, the search methods used, \eg adaptive sampling\new{\cite{blackbox19}}
\sout{\cite{blackbox19, NEURIPS2018_653c579e, wheelerimportance2019}}, bayesian optimization \cite{Abeysirigoonawardena2019Generating}, topic modeling \cite{ding2020learning}, reinforcement learning
\new{\cite{chenbaiming2020}}
\sout{\cite{Kuutti2020, chenbaiming2020, korenadaptive2018}}, flow-based density estimation\cite{multimodaltest} tend to be either highly sensitive to hyper-parameters and proposal distributions\new{\cite{blackbox19}}
\sout{\cite{blackbox19, NEURIPS2018_653c579e, wheelerimportance2019}}or not scale well to high-dimensional search space\new{\cite{Abeysirigoonawardena2019Generating, ding2020learning, chenbaiming2020, multimodaltest}}
\sout{\cite{Abeysirigoonawardena2019Generating, ding2020learning, chenbaiming2020, korenadaptive2018, multimodaltest}}.

Among these, perhaps the closest to our work are evolutionary-based algorithms \cite{interactiontest, pathgeneration, avfuzzer, zhong2021detecting} and their variants (with NN \cite{nsga2nn} or Decision Tree \cite{testing_vision18} for seed filtration) on testing \av or Advanced Driver-Assistance Systems (ADAS).
These methods can scale to high-dimensional input search spaces. Unfortunately, they are currently only used for testing one particular ADAS system or its component (\eg Automated Emergency Braking (AEB) \cite{testing_vision18}, Pedestrian Detection Vision based (PeVi) \cite{nsga2nn},  OpenPilot\cite{zhong2021detecting}, and an integration component \cite{interactiontest}) under one particular \ls, 
testing a controller on road networks without any additional elements (\eg weather, obstacle, and traffic) \cite{pathgeneration}, 
or focusing on finding collision accidents in a \new{\ls}\sout{specific scenario} with other cars constantly changing lanes \cite{avfuzzer}. 
\new{In contrast, our proposed \tool is generalized to different \av systems and scenarios. Our learning-based seed selection and mutation strategy further enables \tool to disclose more unique \bugs than the existing methods. }
\sout{Nevertheless, we}\new{We} adapt the algorithms from \cite{testing_vision18}, \cite{nsga2nn}, \cite{avfuzzer} in our setting, and compare with \tool.

\subsection{Motivating Example}
\label{sec:motiv}

\begin{figure}[ht]
\centering
\includegraphics[width=0.23\columnwidth]{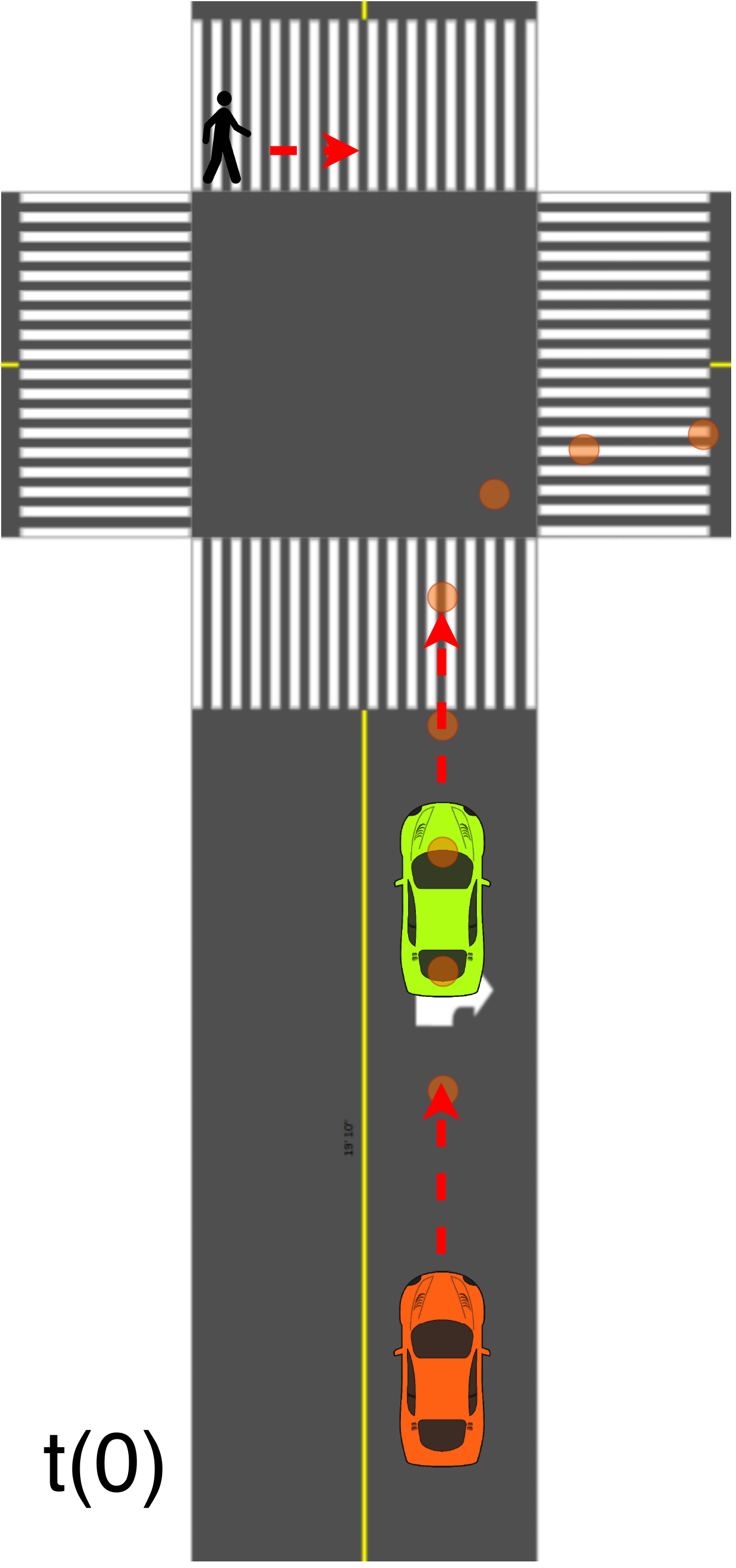}
\includegraphics[width=0.23\columnwidth]{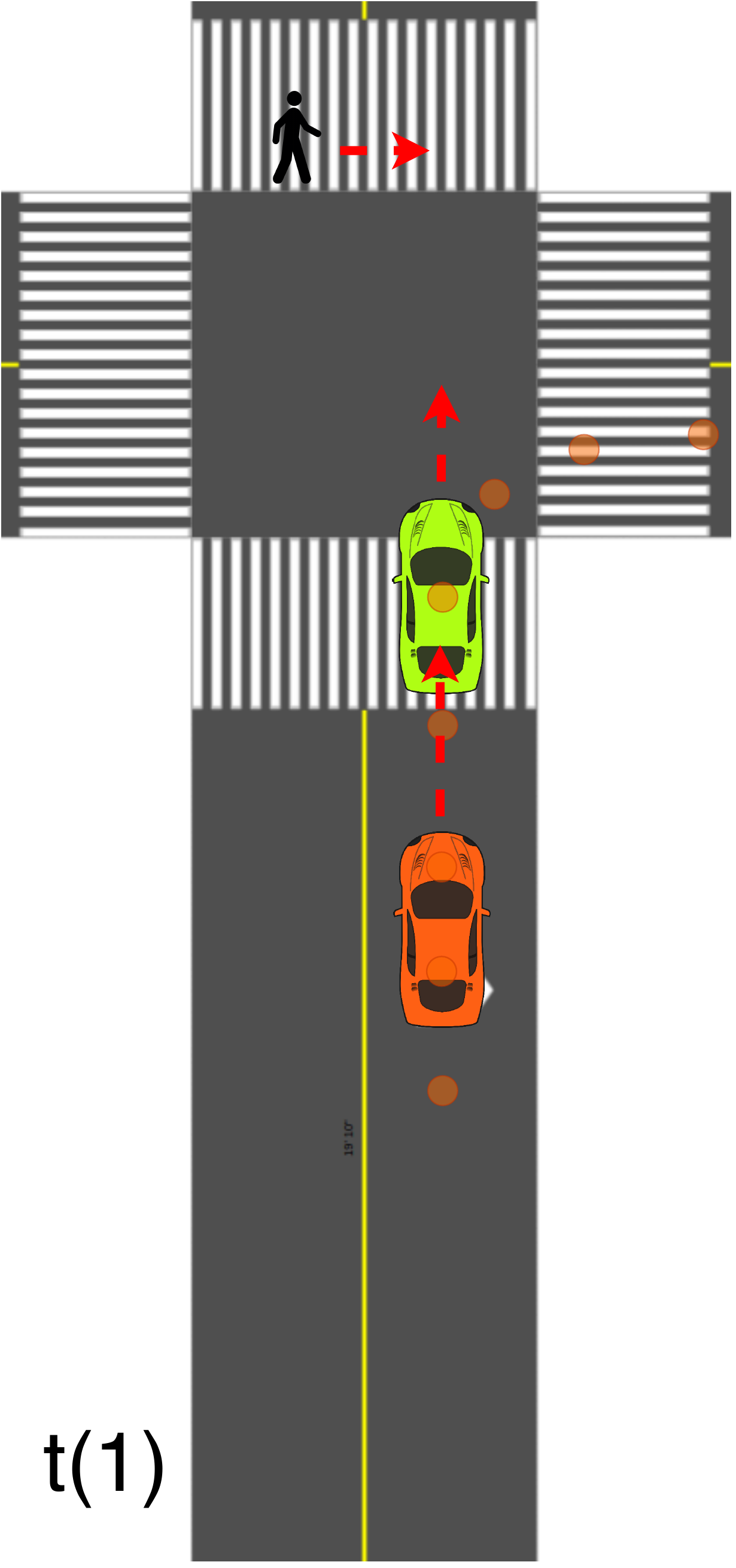}
\includegraphics[width=0.23\columnwidth]{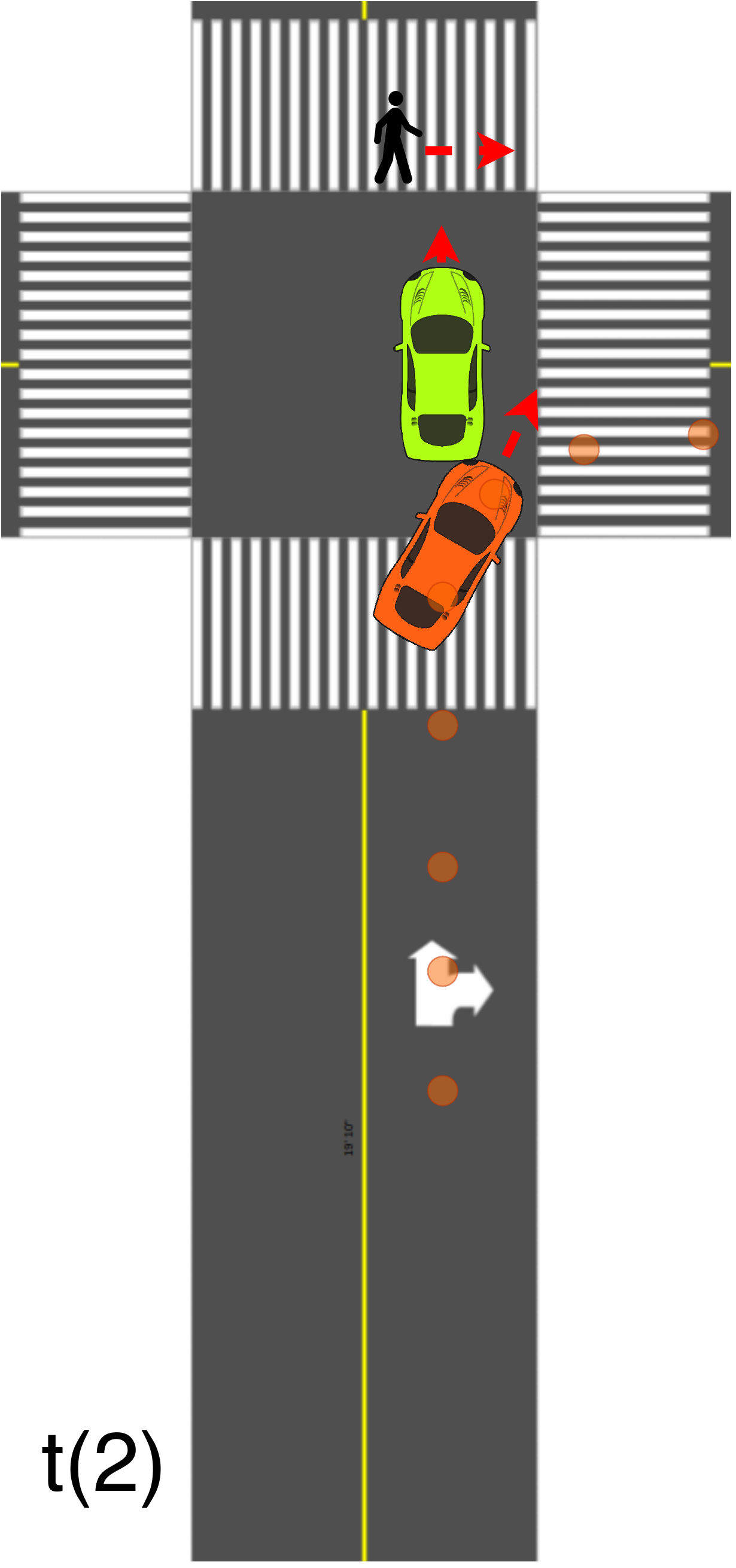}
\caption{\small{\textbf{\edit{Example of Crash Simulation in consecutive time steps.}}}}

\label{fig:scenario_demo}
\end{figure}

\tool aims to generate \bugs by an ego car controller by fuzzing the input scenes. \tool starts with a {logical} driving scenario that involves \bugs, designed based on the top pre-crash \fss from NHSTA  ~\cite{precrash} (see~\Cref{tab:crash}).
For instance, ``vehicle leading ego car stopped'' and ``non-signalized junction" are the top causes of manual car crashes, and \tool tests how an \av behaves in such situations.  ~\Cref{fig:scenario_demo} presents this scenario. 
To simulate a crash in such a situation, \tool starts the simulation with a green car leading an orange ego car near a non-signalized junction (\Cref{fig:scenario_demo}-t(0)). From there, with fuzzing, \tool generates the following crash: the ego-car is going to turn right while the leading car suddenly slows down to avoid hitting a pedestrian who is crossing the road 
(\Cref{fig:scenario_demo}-t(1)). This leads the ego car to collide with the leading car (\Cref{fig:scenario_demo}-t(2)). To simulate the collision, \tool leverages \CARLA's APIs related to vehicle, pedestrian, and cross-road in the map. \newnew{Since the forces that influence collision are mainly the pedestrian’s behavior and the leading vehicle’s behavior,} starting with these agents and starting location in the map, \tool needs to search for valid driving directions 
for all the agents, their speeds, road condition, etc. to simulate the crash. \newnew{Exemplary challenging \specificss in addition to the collision shown here include the pedestrian gets occluded by the leading vehicle (as shown in \Cref{fig:variation_demo}), and the background is at night with heavy rain. The detailed search space of this \ls is provided in Appendix H in supplementary material.}

%% file: body/3_API.tex
\section{API Grammar}
\label{sec:grammar}

\begin{figure}[ht]
\centering
\includegraphics[width=0.9\columnwidth]{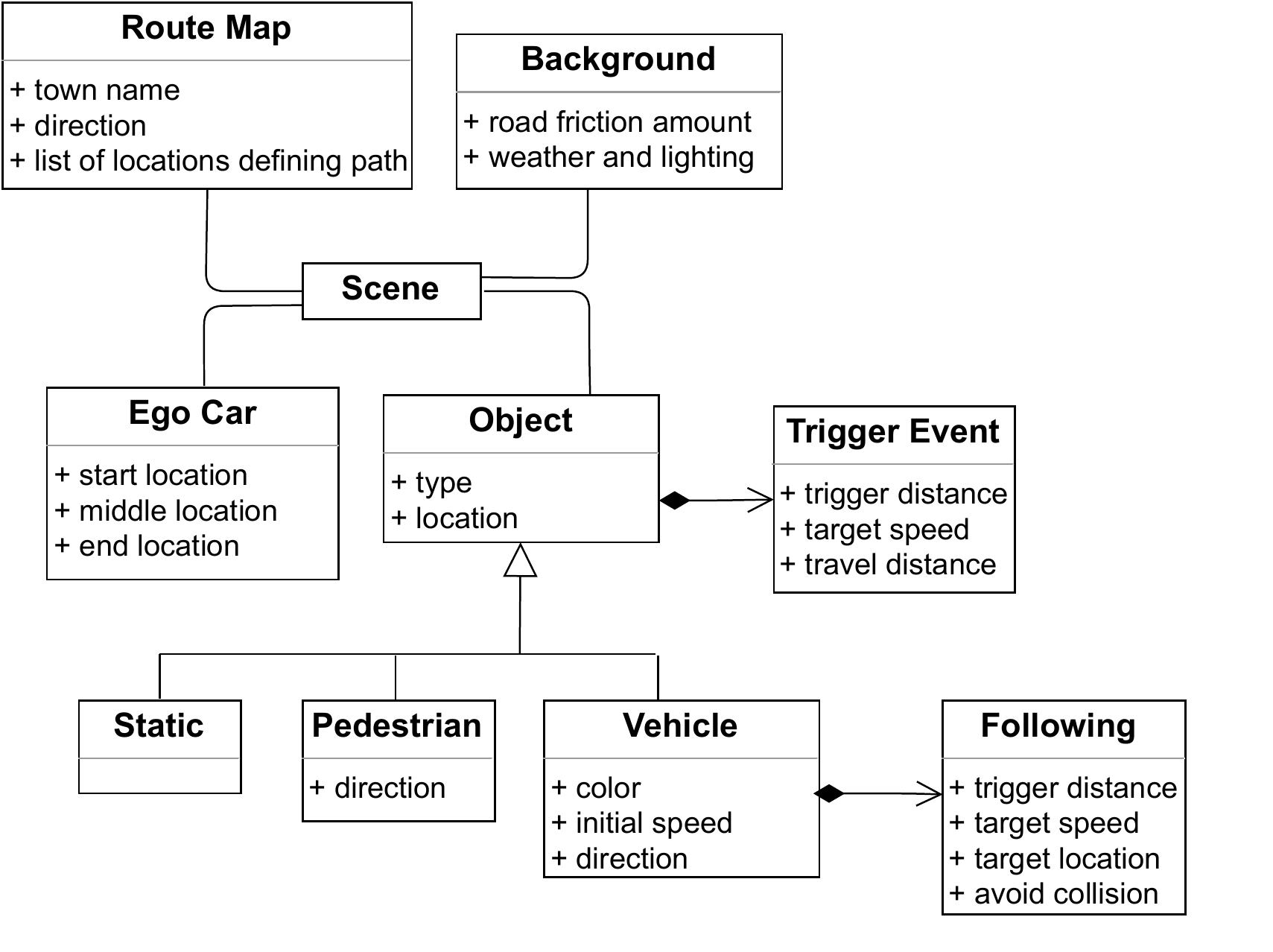}
{
\scriptsize
\begin{tabular}{l|p{6.5cm}}
\toprule   
    \textbf{API} & \textbf{Description} \\  \toprule

    Route Map & The user selects a route map, identified by \textit{town name}, which the ego-car should drive. 
    A map contains a path consisting of a sequence of 2D locations\sout{ defining the route}---the first and last locations in the sequence refer to the start and destination of the ego-car. \CARLA comes with eight predefined maps.
    \\   \midrule
    
    Ego Car &  The controller of the ego car is under test in this paper. 
    \\ \midrule

    Background &  The user can set up a driving environment with different weather and road conditions. 
    The road conditions are set by different friction values. \carla has 21 predefined weather and lighting modes.
    \\ \midrule
    
    Objects & The user can choose a range of static (\eg debris, bus stop, \etc) and moving (\eg vehicles and pedestrians) objects that can appear dynamically around the ego car's route. 
    \sout{Each kind of object can appear in multiple numbers on a route. }Each moving object is associated with a {\em triggering event}, which specifies \sout{when to trigger the object (\ie within a certain distance from the ego-car)}\new{the triggering condition and behavior after being triggered.}\sout{, with what velocity, how long it will travel, and along which direction.} Each vehicle is also associated with a behavior, 
    which makes the vehicle follow CARLA's map with a specified speed to a given destination.
    Users can also choose each vehicle's type (\eg tesla model 3, nissan patrol, \etc), color, and whether to try to drive directly to the destination without regard to other objects.
    \\ 
    
    \bottomrule
  \end{tabular}
}
\caption{\small{\textbf{A simplified description of \CARLA's APIs. We fuzz only over the background and objects.}}}
\label{fig:carla_api}
\end{figure}

Figure~\ref{fig:carla_api} shows a simplified version of the APIs that \tool uses to simulate crashes in our prototype implementation for \carla. The core of the simulation is an initial driving {\em Scene} with four main components: a route map, the ego car whose controller is under test, some static and dynamic objects (\eg other vehicles, pedestrians, \etc), and background like weather and road conditions.


\CARLA 
provides the API specifications as a set of Python APIs~\cite{carla, scenariorunner}. For example, calling \textit{CarlaData\-Provider.request\_new\_actor(pede\-strian\_model, spawn\_point)} creates a pedestrian, where \textit{pedestrian\_model} is a pedestrian asset predefined in \CARLA and \textit{spawn\_point} specifies the pedestrian's initial location and direction. From such specifications we construct a test-generation grammar, $\mathcal{G}$\textit{(Map, Ego Car, Objects, Background)}, shown in Listing~\ref{lst:ped_grammar}. Encoding the grammar in JSON format allows us to specify values for each field. We extend the grammar by adding two constraints for restricting the search region (see Listing~\ref{lst:ped_grammar}) and additional conditions (\eg the distance between the ego-car and the leading car must be greater than a certain distance).

After processing \CARLA's APIs, we get a Test Grammar, $\mathcal{G}$, as $\mathcal{G}$\textit{(Map, Ego Car, Objects, Background, \underline{Search Range}, \underline{Constraint})}, where the underlined 
components\sout{ that facilitate fuzzing} are optional. {The details of search range and constraints are provided in Appendix D in Supplementary Material.} 

\begin{lstlisting}[float,
    caption={An example Test Grammar, $\mathcal{G}$, from \carla's specification. The  \textsc{JSON}-encoded grammar snippet is for the pedestrian in the motivating example. 
    The constraints specified at the bottom express one vehicle's target\_speed $\le 0.5 \times$ of another vehicle's target\_speed}, label={lst:ped_grammar},language=json]
 pedestrian_0: {
  setup: {
    location: {
      x:[-123, -83, (normal, None, 10)], 
      y:[3.5, 43.5, (normal, None, 10)]
    }
    direction: [0, 360],
    type: [0, 12]
  },
  trigger_event: {
    trigger_distance:[2, 50], 
    target_speed: [0, 4],
    travel_distance: [0, 50]
    }}
 
 customized_constraints: [{
   coefficients: [1, -0.5],
   labels: [vehicle[0].trigger_event.target_speed, 
            vehicle[1].trigger_event.target_speed],
   value: 0
   }]

\end{lstlisting}

%% file: body/4_fuzzing.tex
\section{Methodology}
\label{sec:fuzzing}

\gail{this sections keeps saying we do xxx to produce bugs, we do yyy to produce bugs.  I changed some of these to say 'unique bugs' instead of just bugs, but I'm not sure these changes are accurate.  yet the point of this paper is bug diversity, not just large numbers of bugs}

Leveraging the API grammar as described in \cref{sec:grammar}, \tool fuzzes inputs to the ego-car's controller in a black-box manner. 
We make several design decisions to address the following questions:  
(i) How to define {\em unique violation} to simulate {\em diverse} \bugs? (Section~\ref{subsec:unique})
(ii) How to generate only semantically {\em valid} scenes? (Section~\ref{subsec:grammar}) and 
(iii) How to design the fuzzing algorithm to \sout{increase the potential of producing}\new{produce} more {\em valid unique} \bugs? (Section~\ref{subsec:formulation})

\subsection{Diverse Traffic Violations}
\label{subsec:unique}

We focus on two types of violations: collision and going out-of-road. A {\em collision} consists of colliding with other \new{moving}\sout{vehicles, cyclists/motorcycles, pedestrians} or stationary objects. An {\em out-of-road} violation consists of going into a wrong lane (opposite direction traffic), onto the road's shoulder or literally off-road. 


The goal of a good fuzzer should be to find diverse bugs.  However, defining diversity for \bugs is a hard problem. Merely comparing the violation-inducing inputs may lead to infinitely different violations. For example, let's assume that a stationary pedestrian in front of a car results in a crash.  By modifying unrelated input parameters (e.g., the position of another pedestrian far from the crash site, the position of another vehicle in a different lane, etc.), possibly outside the vision of the ego-car controller, we can generate an infinite number of {\em different} violations. But such redundancy is not interesting \new{nor}\sout{or} useful. Thus, criteria for precisely defining \emph{unique} \bugs is needed.
\gail{the intro already says pretty much the same thing has this paragraph}
\ziyuan{It expands the intro a bit. If no space left, we can potentially make this para shorter.}


Abdessalem \etal~\cite{testing_vision18} define that two test {\specificss} are distinct if they differ in "the value of at least one static variable or in the value of at least one dynamic variable with a significant margin." This definition fails in our high-dimensional scenarios, 
as the example above could be considered different violations by their criteria. We instead count the number of unique violations as:



\noindent
\textbf{Unique Violation.} {For a given type of \bug (collision or out-of-road), two violations caused by {\specificss} 
x and y are \emph{unique} 
if at least $th_1\%$ of the total number of changeable fields 
are different between the two, where $th_1$ is a configurable threshold.}

For a discrete field, the corresponding values are different if they are non-identical in $x$ and $y$ (\eg ``color" field is different between a black and a white car). 
For a continuous field, the corresponding normalized values should be distinguishable by at least $th_2\%$, where  $th_2$ is a user-defined threshold. For instance, if the speed range of a car is $[0,10]$m/s, and two violations occur at speeds 3m/s and 4m/s, the field is considered to be the same between the two violations since $\textstyle \frac{4-3}{10-0}=0.1<0.15$, where $th_2\%=15\%$.

\noindent\new{\textbf{Scalability of the Definition.} Compared with the definition in \cite{testing_vision18}, the new definition has two benefits in terms promoting the violation's diversity for higher dimensional logical scenarios. First, since $th_1\%$ is the percentage of the total number of changeable fields that need to be different, given a fixed $th_1\%$, as the number of changeable fields goes up, the number of changeable fields that need to be different for two violations to be considered different also goes up. Second, the current definition enables a user to specify the thresholds $th_1\%$ and $th_2\%$ according to one’s need. For scenarios with higher dimensions, larger $th_1\%$ can be used.}

\noindent\textbf{Benefits of Finding More Unique Violations.} There are two benefits of finding more unique violations for AV testing. First, it enables engineers to better identify the limitation of the AV under test. Different violations can potentially expose different functionality issues and/or with different causes. \newnew{Compared with the formulation of finding the Pareto front as in \cite{nsga2nn}, our method allows more exploration and thus can find not only the most severe violations but also less severe violations that should be avoided and can be potentially useful for improving the AV under test (e.g., collisions at low speed). Such violations tend to be missed by methods optimized for Pareto front since they usually generate new seeds based on the most extreme violations so far at each generation.} Second, by maximizing the number of unique violations found, the “violation coverage” in the user specified \ls is maximized. Instead of maximizing the “branch coverage” as in the traditional fuzz testing, we maximize the number of unique \specificss (for each \ls) that induce violations. This can help a tester to validate if an AV can perform well in the specified \ls as expected.

\subsection{Fuzzing with API Grammar}
\label{subsec:grammar}

\tool takes the API grammar as input and fuzzes following the grammar spec. The user first selects a route map where the ego-car controller will drive and a starting \is\sout{, which is} encoded according to the API grammar. Users can optionally specify a \edit{customized} search region and constraints. \tool uses these pieces of information to sample \edit{\iss (also called seeds in \sout{the }fuzzing\sout{ context})};
Each sampled \edit{\is} obeys the constraints enforced by the API grammar. 



Figure~\ref{fig:workflow} shows a high-level overview of the fuzzing process. The objective is to search for \iss that will lead to unique \bugs. To achieve this, like common blackbox fuzzers, 
\gail{this paper says too many times that we do the same thing as everyone else}
\tool runs iteratively: \tool samples the grammatically valid \edit{\iss} (Step-1), and the simulator runs these \edit{\iss} with the controller under test to collect the results as per the objective functions, as detailed in ~\Cref{subsec:obj}. \tool leverages feedback from previous runs to generate new seeds, \ie favors the ones that have better potential to lead to violations over others (Step-II) and further mutates them (Step-III).  The API grammar constraints are followed while incorporating feedback to create new mutants, so all the mutants are also semantically valid. The new seeds are then fed into the simulator to run. The \bugs found are reported, and their corresponding seeds added to the seed pool. This repeats until the budget expires. 




\subsection{Fuzzing under Evolutionary Framework}
\label{subsec:formulation}

\tool aims to maximize the number of {\em unique} \bugs found within a given resource budget (\eg \# simulations). This is 
an optimization problem, where \tool searches over the entire input space of grammatically valid initial scenes to maximize unique violations found by simulating from those scenes. More formally, if \allx is the space of all possible valid input scenes, \tool searches over \allx to maximize \bug count (\ally) within a fixed budget, say $\mathcal{T}$.  Thus, if $B_t$ is the set of \bugs found by input $x_t\in\mathcal{X}$ at fuzzing step $t$, 
then more 
formally fuzzing is: $\scriptstyle{\mathcal{Y_\mathcal{T}} = \max\norm{~\bigcup_{t=1}^\mathcal{T} B_t}}$. Here $\scriptstyle \|\cdot\|$ is the norm and $\scriptstyle{\bigcup(\cdot)}$ represents the union of all violations over all possible inputs.

Since the input space \allx is prohibitively large, an exhaustive search to optimize the equation is infeasible. Instead, one needs to identify and focus the search on promising regions to optimize the number of unique violations. Fuzzing based on evolutionary algorithms is a common approach 
\gail{another place where we say we do what everyone else does}
for such optimization. Starting with some initial inputs, evolutionary fuzzers tend to select new inputs that find new violations and further mutate those successful inputs to generate further new inputs.  Thus, the success of fuzzing depends on careful design of the following three parts:

\squishlist
\item[(i)] \textit{Objective function ($F$)}: How to design a objective function 
to maximize unique bugs? 
\item[(ii)] \textit{Seed Selection ($x\in$\allx)}: Which inputs to mutate~\cite{bohme2017coverage}?  and
\item[(iii)] \textit{Mutation($m$)}: How to mutate~\new{\cite{lemieux2017fairfuzz, chen2018angora, she2018neuzz}}\sout{\cite{lemieux2017fairfuzz, vuzzer,chen2018angora, she2018neuzz}}?   
\squishend



Thus, the next generated input at time t, $x_{t}$ depends on ($x_{:t-1},m$), where $x_{:t-1}:=x_1,...,x_{t-1}$.
The set of \bugs $B_t$ found by $x_t$ can be represented as a function ($F$) of these fuzzing parameters, \ie ~$B_t=-F(x_{:t-1},m)$, such that minimizing $F$ will maximize the unique \bugs. 
Thus, more formally, evolutionary fuzzing (with $x_0$ is an initial seed input) can be written as: 
{
\setlength{\abovedisplayskip}{0pt}
\setlength{\belowdisplayskip}{0pt}
\begin{equation}
    \label{eq:formal_main}
    \scriptstyle
    \mathcal{Y_\mathcal{T}} 
    = \min_{x_{:t-1},m}\norm{~\bigcup_{t=1}^\mathcal{T}F(x_{:t-1},m)~}
\end{equation}
\vspace{-3mm}
}

\edit{In the following, we discuss the details of the fuzzing. 
}

\subsubsection{Objective Function.}
\label{subsec:obj}

The ultimate goal of the fuzzing algorithm is to maximize diverse \bugs found. However, as the bug-producing inputs are sparse, we need more violation-specific guidance to help the ego car move towards the violation points. For example, to generate a collision with a pedestrian, we need to guide both the ego car and the pedestrian closer to each other. Thus, we need a {\em smoother} objective function that helps lead towards the \bug. To this end we define the following objective functions: 
\gail{again, we say this work has already been done}
\ziyuan{I removed "adapted from previous work... since we mentioned later}



{
\setlength{\tabcolsep}{3pt}
\renewcommand{\arraystretch}{1}
\vspace{1mm}
\centering
\scriptsize
\begin{tabular}{lll}
\toprule
\textbf{Violation} &  &  \\ 
\textbf{Type} & \textbf{Objective}  & \textbf{Definition} \\ \toprule
 \multirow{3}{*}{Collision} &   $F_{collision}$ &:= \textrm{speed of ego-car at collision} \\
    & $F_{object}$ &:= \textrm{minimum distance to other objects} \\ 
    & $F_{view}$ &:= \textrm{minimum angle from camera's view} \\ \midrule
\multirow{3}{*}{Out-of-road}    & $F_{wronglane}$ &:= \textrm{minimum distance to an opposite lane} \\
    & $F_{offroad}$ &:= \textrm{minimum distance to a non-drivable region} \\
    & $F_{deviation}$ &:= \textrm{maximum deviation from interpolated route} \\ 
\midrule
\end{tabular}
\vspace{1mm}
}


\noindent
\textit{Collision.} We optimize for the weighted sum of the three smooth objective functions: $F_{collision}$, $F_{object}$, and $F_{view}$, {similar to the objectives used in~\cite{testing_vision18, nsga2nn, interactiontest}}. $F_{collision}$ and $F_{object}$  promote the severity of collision and the chance of collision, respectively. $F_{collision}$ is set to $-1$ as per \cite{testing_vision18} when no collision happens. $F_{view}$ promotes cases where the object(s) involved are within the camera(s) view.
\gail{again we do the same as cite cite cite}



\noindent
\textit{Out-of-road.} This is implemented by a weighted sum of the three smooth objectives: $F_{wronglane}$, $F_{offroad}$, and $F_{deviation}$. $F_{deviation}$ is adapted from the objective of "maximum distance deviated from lane center" in \cite{Kuutti2020}. 

We further define $F_{wronglane}$ and $F_{offroad}$ to strengthen the signals for driving into an incorrect lane or off the road, respectively. \edit{\Cref{fig:wronglane_objective} in \Cref{sec:illustration_of_wronglane_objective} provides an illustration.}


\begin{figure}[h]
 \centering
 \includegraphics[width=0.48\textwidth]{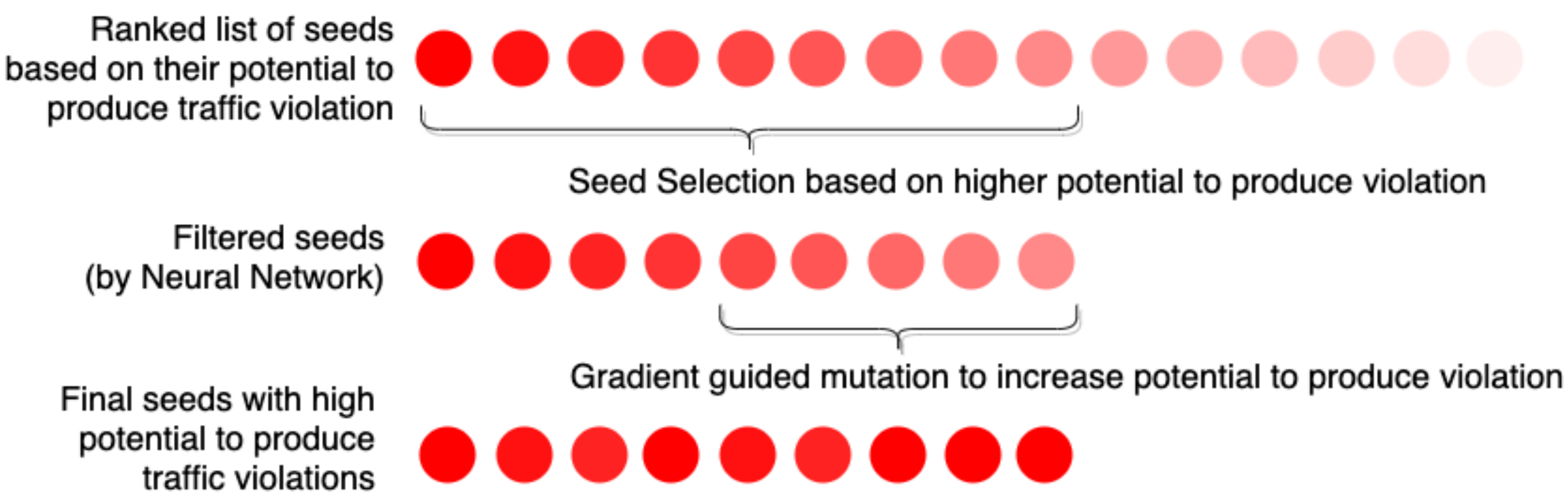}
 \caption{Seed Selection \& Mutation Strategy per Generation}
 \label{fig:mutation}
\end{figure}


For each \bug type, we formulate the fuzzing problem as a constrained multi-objective optimization\sout{ problem}. Let $x$ be an input, \ie a \new{\specifics}\sout{\specificss} with all the searchable fields. Denote $F_i(x)$ for $i=1,...,n$ to be $n$ objective functions, $w_i$ to be some user-provided weights, and $g_j(x)$ for $j=1,...,p$ to be $p$ constraints, where each constraint is expressed as $\leq 0$ form. Then, the objective function $F(x)$ of ~\Cref{eq:formal_main} can be expressed as a constrained weighted sum:  
$\textstyle   \textrm{min}_x~\sum_{i=1}^{n} w_i F_i(x), 
    \textrm{s.t. } g_j(x) \leq 0 ~\forall j=1,...,p$.
Unlike \cite{testing_vision18, nsga2nn}, we optimize for a weighted sum of objective functions rather than search for a Pareto front of the involved objective functions, because our goal is to find the maximum number of unique \bugs rather than \bugs with the Pareto front of multiple objectives. 

\subsubsection{Seed Selection.}
\label{subsec:seed_selection}

Common evolutionary fuzzers like AFL~\cite{zalewski2017american} maintain a seed queue and tend to favor some seeds over others. Smart seed selection strategies give a significant boost to fuzzing performance to not waste limited resources by running fruitless seeds~\cite{yue2020ecofuzz, bohme2020boosting}. In our case, a bad seed may lead to running several scenes without simulating a \bug. We devise an incremental learning-based seed selection strategy, as shown in~\Cref{fig:workflow}. 
\gail{so much of this paper sounds like we're saying "we do the same thing as everyone else"}


For each generation $t$ of our evolutionary search, a Neural Network ($NN_{t-1}$) is trained with all the seeds executed up to generation $t-1$, such that the NN learns to differentiate between successful vs. unsuccessful seeds. $NN_{t-1}$ is used to predict the seeds generated in generation $t$. It ranks all the candidate seeds of generation $t$ based on its confidence of leading to a unique \bug. \tool then selects the top $S$ seeds that are more likely to produce violations, where $S$ is a configurable parameter. Figure~\ref{fig:mutation} illustrates this process. The top row shows all the seeds generated in a particular generation. The NN ranks them based on their potential to produce unique violations\textemdash darker color is more violation prone than lighter. The top $S$ seeds are then selected for future steps (in the second row.)


\subsubsection{Mutation.}
\label{subsec:mutation}

Among the top $s$ seeds selected in the previous step, not all are equally likely to lead to unique violations. In particular, the NN has lower confidence on the bottom seeds of the ranked list (the lighter color seeds in the second row of Figure~\ref{fig:mutation}). \tool further mutates such lower confidence seeds to increase their potential to simulate \bugs. A constrained gradient-guided perturbation mutates the lower confidence seeds towards higher confidence (the third row in Figure~\ref{fig:mutation} where all the seeds become dark red). This perturbation is generated by iteratively back-propagating the input's gradient with respect to the NN's prediction. We describe the perturbation algorithm in ~\Cref{sec:step3}.

%% file: body/5_implementation.tex
\section{Implementation Details}
\label{sec:implementation}

We realize our evolutionary fuzzing design discussed in ~\Cref{sec:fuzzing} following the main steps: Sampling, Seed Selection, and Mutation (see Figure~\ref{fig:workflow}). 
{Appendix A - Algorithm~\ref{alg:ga-un-dnn-adv} in Supplementary Material gives the detailed algorithm.} 

\paragraph*{\textbf{Step-I: Sampling.}}
\label{sec:step1}

This step samples seed test cases from the entire input space by obeying the constraints enforced by the API grammar. 
We use two sampling strategies: (i) random and (ii) genetic algorithm (GA). Each field is sampled based on a user-specified distribution, search range, and constraints (see Listing~\ref{lst:ped_grammar}). 
In either strategy, when the specified constraints are not satisfied, each variable will be re-sampled. If the specified constraints \sout{have very small probabilities }and cannot be satisfied after a specified number of attempts, the program will raise an error.  
We filter out seeds similar to those corresponding to previous relevant \bugs. In the fuzzing literature, this step is commonly used for \textit{test suite minimization}~\cite{interactiontest}.


At each generation, the GA considers the previous seeds with results, selects from them new parent test cases, and generates new seeds through crossover and mutation. 

\noindent \textit{Selection}: 
We adopt binary tournament selection with replacement, like the original \NSGA implementation~\cite{nsga2}, \edit{as well as the variations in \cite{testing_vision18, nsga2nn}}. 
Two duplicates are created for each sample and randomly paired. Each pair's winner is then randomly paired as the parents for this generation's mating process. The rank of two individuals is determined by the objective function in \Cref{subsec:obj}. 


\noindent \textit{Crossover \& Mutation}: 
Simulated Binary Crossover \cite{SBX}, a classical crossover method commonly used for floating point numbers, 
is adopted\edit{, as in \cite{nsga2, testing_vision18, nsga2nn}}. 
A distribution index ($\eta$) is used to control the similarity of the offspring and their parents. The larger $\eta$ is, the more similar the offspring are \wrt their parents. We set $\eta=5$  and probability=$0.8$ to enable more diversity. \edit{If a larger $\eta$ is used, the offspring will be more similar to their parents, so it takes longer to find distinct offspring for methods with uniqueness filtration and results in fewer unique bugs found for methods without. If a smaller $\eta$ is used, the offspring will be too distinct from their parents and violation-inducing parents won't be fully leveraged.} Polynomial Mutation is applied to each discrete and continuous variable~\cite{polymutation}. For discrete variables, we treat the value as continuous during the mutation and round later. We clip the values at specified boundary values. Following \cite{testing_vision18}, mutation rate is set to $\frac{5}{k}$, where $k$ is the number of variables per instance. We further set the mutation magnitude $\eta_m$ to $5$ for larger mutations.

\paragraph*{\textbf{Step-II: Seed Selection.}}
\label{sec:step2}
As described in ~\Cref{subsec:seed_selection}, we boost fuzzing performance with a learning-based seed selection strategy. 
We train a shallow neural network (1-hidden layer) using the previous seed test cases to predict if a test case leads to a \bug. The NN ranks the next generation seeds based on its confidence of leading to a \bug and the most likely tests are selected.
\gail{any bug? or a unique bug?}
\ziyuan{The uniqueness thing is guaranteed by the filtration step so the selected seeds are already distinct enough. This selection using NN by itself does not optimize for uniqueness but it happens after the filtration.}
Some previous work \cite{nsga2nn} also leverages an NN for seed selection. 
There are several major differences. First, we train a single NN for binary classification of \bugs rather than several NNs for regressing over all objective values as in~\cite{nsga2nn}. Thus we rank test cases based on the confidence value of finding a \bug rather than the Pareto front from multiple NNs. This design choice is motivated by our goal to find maximum number of valid, diverse \bugs rather than finding \edit{the best set of} \bugs achieving \edit{the optimal trade-off among multiple objectives at the same time}. 
\gail{the way this is phrased, the other one sounds better}
Second, we iteratively train the NN in an active learning setting rather than training fixed ones at the beginning. This active training results in increasingly more training samples than the initial population and, thus, improved NN approximation over time. We show both design choices introduce performance gains in the experiment section. 


\paragraph*{\textbf{Step-III: Constrained Gradient-Guided Mutation.}}
\label{sec:step3}
As per ~\Cref{subsec:mutation}, we apply a constrained gradient-guided mutation 
on the selected top test cases to maximize their likelihood of leading to \bugs. 
\gail{any bugs or unique bugs?}
\ziyuan{The uniqueness thing is guaranteed by the filtration step and line15 in Algorithm1 which will stop the mutation if the next mutation makes it similar to previous bug but the mutation by itself does not optimize for uniqueness.}
The procedure, shown in Algorithm~\ref{alg:attack}, is adapted from the constrained adversarial attack in \cite{constrainedpgd}. 
\begin{algorithm}
    \scriptsize
    \SetKwInOut{Input}{Input}
    \SetKwInOut{Output}{Output}
    
    \Input{$\mathbf{x}$: test case, $\mathbf{f}$: NN forward function of predicting a test case's likelihood of being a \bug, $\mathbf{th_{conf1}}$: threshold of conducting a perturbation, $\mathbf{th_{conf2}}$: threshold of stopping a perturbation, $\mathbf{n}$: maximum number of iterations, $\mathbf{\lambda}$: step size, $\mathbf{c}$: constraints, $\mathbf{\epsilon}$: maximum perturbation bound, $\mathbf{x_{min}}$: minimum allowable\sout{ input} values, $\mathbf{x_{max}}$: maximum allowable\sout{ input} values}
    \Output{$\mathbf{x'}$: mutated test cases}
    
    $x' = x$\;
    $i = 0$\;
    \If {$f(x)>th_{conf1}$}{
        return $x$\;
    }
    
    \While {$i < n$}{
        $i+= 1$\;
        $dx$ = $\lambda \frac{df(x')}{dx'}$\;
        $x' = x' + dx$\;
        $x' = clip(x', x_{min}, x_{max})$\;
        $dx = clip(x' - x, -\epsilon, \epsilon)$\;
        
        \If {check-constraint-violation ($c$, $dx$) == True}{
            $dx$ = linear-regression ($c$, $dx$)\;
        }
        \If {is-similar ($X$, $x+dx$)}{
            break\;
        }
        $x' = x + dx$\;
        
        \If {$f(x') > th_{conf2}$}{
            break\;
        }
    }
    
    \Return $x'$

    \caption{\textbf{\small{Constrained Gradient Guided Mutation}}}
    \label{alg:attack}
\end{algorithm}
A test case $x$ is perturbed only when the NN's confidence in its leading to a \bug, $f(x)$, is smaller than a threshold $th_{conf1}$. If a test case is already considered highly likely to lead to a \bug, there may be no extra benefit in further perturbing it. Otherwise, an iterative process begins (line 6-21). At each iteration, a small perturbation $dx$ is generated (line 8) via back-propagation from maximizing the test case's NN confidence. The perturbation is then clipped based on allowable input value domains and a user-specified maximum perturbation bound $\epsilon$ (line 9-11). Next, the perturbation is checked against grammar constraints (line 12). If necessary, a linear regression projects it back within the constraints. The perturbed test case is then checked against previously found \bugs (line 15). If a similar test case already found a \bug, the perturbation process ends, and the latest perturbation won't be applied. Otherwise, the current perturbation is applied on top of the perturbed test case from the last iteration (line 18). The new perturbed test case is then fed into NN for its confidence of leading a \bug. If larger than a specified threshold $th_{conf2}$, the mutated test case will be returned and the mutation procedure ends. Otherwise, a new iteration begins.


\noindent
\textbf{Enforcing Grammar during Feedback.} One difficulty here is to make sure the perturbed test case still satisfies the grammar constraints. The simplest solution is to discard the perturbations (and subsequent iterations) that lead to constraint violation. However, as shown in ~\cite{constrainedpgd}, the insight for linear constraints is if an original (unperturbed) test case satisfies the constraints and the perturbation alone satisfies the constraints as well, then the perturbed test case also satisfies the constraints. Thus, only the perturbation needs to be checked against the constraints after each iteration. If some constraints are violated, we apply a linear regression to the perturbation to map it back within the constrained region (motivated by~\cite{constrainedpgd}). For the linear regression, the non-constant part of the constraints are weights $\mathbf{W}$ where each row corresponds to the coefficients of one constraint, the constant parts $y$ are the objectives, and the projected perturbation $dx_{proj}$ are the variables to search for. The linear regression starts with the perturbation $dx$ and find the the projection $dx_{proj}~=~ \textrm{arg\,min}_{ dx_{proj} }\norm{\mathbf{W} dx_{proj}~-~y}$.

%% file: body/6_experiments.tex
\section{Experimental Design}
\label{sec:experiment-design}

\noindent
\textbf{Environment.} Our primary evaluation uses the \CARLA \new{(}version 0.9.9\new{)} simulator \cite{carla}. \new{To show the generalization of our approach, we further conduct evaluation using the \svl (version 2021.3) simulator \cite{svl} in RQ4.} All the algorithms are built on top of pymoo~\cite{pymoo}, an open-source Python framework for single- and multi-objective algorithms.


\noindent
\textbf{Scenarios.} We run \tool under five different \lss \new{(\Cref{tab:experiment_scenarios})} inspired by the NHTSA report~\cite{precrash}.\sout{The details are shown in ~\Cref{tab:experiment_scenarios}.}

\noindent\new{\textit{Selection.}} The first three \lss cover the top six pre-crash \fss \new{in terms of frequency and incurred economic cost as} shown in Table \ref{tab:crash}\sout{, and the fourth also occurs frequently. The fifth is also a common \ls}. \new{The last two scenarios are also common \lss ranked fourth and eighth among the pre-crash scenarios of two-vehicle light-vehicle crashes in terms of occurrence frequency~\cite{precrash}. Note that the other frequent scenarios have been covered by the first three.}

\noindent\new{\textit{Adaptation.} To use a NHTSA pre-crash scenario for testing, we let the ego car be a vehicle involved in each crash scenario. To make the scenario searchable, we convert each functional scenario into a logical scenario (defined in \Cref{sec:definition}) that satisfies the functional scenario’s description. For example, in the scenario “A leading vehicle decelerating/stopped”, the ego car is the following vehicle. We set the search range of the location of the leading vehicle to be in the same lane and ahead of the ego car. Additionally, we set the search range of the speed of the leading vehicle after being activated to be slower than that of its initial speed. These designs enable every generated specific scenario to satisfy the logical scenario “A leading vehicle decelerating/stopped”.}

\noindent\new{\textit{Validity.} The scenarios we use are supposed to be within the operation design domain (ODD) of the \av controllers under test which are all supposed to handle regular traffic scenarios. To check that they can handle the base scenarios, we conduct a validity test for each scenario. The result shows that, when no other vehicles/pedestrians are present, the corresponding AV controller can successfully reach its destination without incurring \bugs. When there are other vehicles/pedestrians, the corresponding controller succeeds with no violations in some cases but not others. Our goal is to find those violations.}



{
\setlength{\tabcolsep}{3pt}
\renewcommand{\arraystretch}{1}
\begin{table*}[th]
    \centering
    \scriptsize
    \caption{\textbf{\small{Different Driving scenarios under Test}}}
    \label{tab:experiment_scenarios}
    \setlength{\tabcolsep}{3pt}
    \renewcommand{\arraystretch}{1}
    \begin{tabular}{l|l|l|r|l|r}
    \toprule
                   & Corresponding  &    &   & Road  & \#violations \\
        Logical Scenarios Names  & NHTSA functional scenarios* &  \#Para  &  Map ID & Type & found**\\
    \toprule
    Turning right while leading car slows down/stops  & Leading vehicle stopped / deccelerating & 26 & town05 & junction & 512 \\
    Turning left a non-signalized junction & Vehicle(s) turning at non-signalized junctions & 26 & town01 & non-signalized T-junction & 672 \\
    Crossing a non-signalized junction & Straight crossing paths at non-signalized junctions & 47 & town07 & non-signalized junction & 400 \\
    Changing lane & \new{Vehicle(s) changing lanes – same direction}\sout{n/a} & 26 &  town03 & straight road & 147 \\
    \newedit{Turning left a signalized junction} & \new{LTAP/OD at signalized junctions}\sout{n/a} & 11 & Borregas & signalized & 76 \\
    \bottomrule
    \end{tabular}
    
    *all scenarios involve ego car lost control or drove off-road, without taking any action, by testing if the ego-car goes out-of-road.\\
    ** (first four rows) average numbers of collision \bugs (for town03 and town05) or out-of-road \bugs (for town01 and town07) found by \GAUNNNGRAD on the \textbf{lbc} controller in \carla. (last row) average number of collision \bugs found by \GAUNNNGRAD on \apollo in \svl.
    
\end{table*}
}

\noindent
\textbf{\av controller.} 
\newedit{We test two rule-based PID controllers, \textbf{pid-1} \cite{ding2020learning} and \textbf{pid-2} \cite{chen2019lbc}, one end-to-end controller \cite{chen2019lbc}, (\textbf{lbc}), and one modular controller \cite{apollo}, (\textbf{\apollo}). 
lbc is a vision-based, end-to-end controller proposed in \cite{chen2019lbc}.
PID controllers assume knowledge of the states of other objects in the environment and the trajectory to follow. They attempt to reach the next planned location with a specified speed by adjusting controls for brake, throttle, steering and try to minimize the mismatch with the desired speed and direction while avoiding collision with other objects. 
pid-1 is a default rule-based controller in CARLA’s official release \cite{carla} and has been used as the main system under test in existing literature \cite{ding2020learning}. pid-2 is a rule-based pid controller implemented by the authors in \cite{chen2019lbc} to collect data to train lbc. 
\apollo is an industrial-grade, modular controller\cite{apollo}.}

\noindent
\textbf{Hyper-parameters.}
The NN for seed selection has a hidden layer of size 150. We use the Adam optimizer with 30 epochs and batch-size 200. $th_{conf1}$ is set to be the $0.25 \times p$-th highest NN confidence value among training data, where p is the percentage of the training data leading to \bugs, and $th_{conf2}$ is set to 0.9. $\epsilon$ is set to be $1$, $n$ is set to 255 and $\lambda$ is set to $1/255$ so an input seed can be perturbed to any other input seed in the input domain. We collected seeds up to 10 generations (and thus 500 simulations) by default. The default method used for seed collection is \GAUN. 

\noindent
\textbf{Metrics.}
When we compare search quality, we use the number of \emph{unique} \bugs found over the corresponding number of simulations run. We use the number of simulations rather than time because the former is platform independent. Moreover, the time costs mainly come from simulations. On average, each simulation takes about 40 seconds, while the generation process only takes about 10 seconds and is only invoked once per generation. A simulation ends if a violation happens, the ego car reaches the destination, or time (50 seconds) runs out. 
When counting collision \bugs, for lbc, pid-1, and pid-2, we only count those where the collision happened within the view of the controller’s front camera and the controller did not stop to avoid the collision. For \apollo, since it is equipped with LiDAR (providing 360 degrees view), we count all collision \bugs where it did not stop to avoid them. We further manually checked a set of found collision scenarios and found they can be avoided if the controllers maneuver correctly. For example, in \Cref{fig:more_bugs_demo_apollo}, if \apollo slows down earlier, both collisions could be avoided.
\newnew{When the baseline method \AVFUZZER is considered (i.e., \Cref{fig:search_methods_on_different_scenes} and \Cref{fig:apollo_num_collision_found}), since it does not have a seed collection stage, for a fair comparison, the number of simulations for the seed collection stage of other methods is also included. When the comparison does not involve \AVFUZZER, the number of simulations for seed collection is excluded since all the methods will be set to share the same seed collection stage for a fair comparison.}\remove{The number of simulations referred to in the RQ1 and RQ2 excludes the number of simulations needed for the seed collection stage since all the methods will be set to have the same seed collection stage for a fair comparison. In RQ4, since the baseline method \AVFUZZER does not have the seed collection stage, for a fair comparison, the number of simulations includes the seed collection stage for the other methods.}
We set uniqueness thresholds $th_1=10\%$ and $th_2=50\%$ as default values, and explore the sensitivity of different search methods under nine different combinations.

\begin{table}[h]
\vspace{0.2cm}
    \scriptsize
    \centering
    \caption{\textbf{\small{Proposed methods, baselines and variations}}}
    \label{tab:all_methods}

    \begin{tabular}{l|l}
    \toprule
        Method & Description  \\
    \toprule
        \tool (\GAUNNNGRAD)  & \GAUNNN w/ constrained gradient guided \\
        & mutation  \\ 
        ($\epsilon$=1.0) & \\
        \midrule
        \textbf{Baselines} & \\ 
        \NSGADT \cite{testing_vision18} & \NSGA w/ decision tree  \\
        \NSGASM \cite{nsga2nn} & \NSGA w/ surrogate model   \\
        \NSGAUNSMA  & \NSGASM w/ duplicate elimination and   \\
        & incrementally learned surrogate model \\
        \newedit{\AVFUZZER} \cite{avfuzzer} & \newedit{global \GA + local \GA} \\
        
        \midrule
        
        \textbf{Variants} & \\ 
        \GAUNNNGRAD * ($\epsilon$=0.3) & \GAUNNNGRAD w/ a smaller (0.3 rather   \\
         & than 1) maximum perturbation bound $\epsilon$ \\
        \RAUNNNGRAD & \RA w/ duplicate elimination,   \\
        & NN filtration and constrained gradient \\
        & guided mutation \\
        \GAUNNN & \GAUN w/ NN filtration   \\
        \GAUN  & \GA w/ duplicate elimination   \\
        \GA  & genetic algorithm  \\
        \RA & random sampling  \\
    \bottomrule
    \end{tabular}
    
    \scriptsize{* GA= Genetic Algorithm, UN = Unique, NN = Neural Network based seed selection, GRAD=Gradient guided mutation}
\end{table}

\noindent
\textbf{Baseline Comparison.}
We compare \tool with three baseline methods shown in \Cref{tab:all_methods}'s baselines row. To fairly compare the fuzzing strategies on equal footing, we used the same objectives from \Cref{subsec:obj} and the same random sampling with uniqueness filtration to generate the initial populations for all. We also compare \tool with alternative design choices in \Cref{tab:all_methods}'s variants row. 

\newedit{Among the baseline methods, \NSGADT and \NSGASM are two multi-objective GA-based methods and \AVFUZZER is a single-objective GA-based method, all of them are adapted from previous work \cite{testing_vision18, nsga2nn, avfuzzer}.}
\NSGADT calls \NSGA \cite{nsga2} as a subroutine. After each run of \NSGA, \NSGADT fits a decision tree over all instances so far. It uses cases that fall into the leaves with more \bugs than normal cases (a.k.a. "critical regions") as the initial population for \NSGA's next run. During \NSGA, only the generated cases that fall into the critical regions are run. We set search iterations to 5 as in \cite{testing_vision18}. Since the tree tends to stop splitting very early in our \edit{\lss}, we decrease the impurity split ratio from 0.01 to 0.0001. We set minimum samples split ratio set to $10\%$.

\NSGASM trains regression NNs for every search objective and ranks candidate test cases and \bugs found so far based on the largest Pareto front and crowding distance, as in \NSGA. \edit{To further compute the effects of uniqueness and incremental learning as well as the effects of weighted sum objective and gradient-guided mutation, we implement \NSGAUNSMA --- a variant of \NSGASM with additional duplication elimination and incremental learning.} \edit{For both \NSGASM and \NSGAUNSMA training processes, we first sampled 1000 additional seeds to train three regression NNs. For finding collision violations, the three NNs are trained to predict $F_{object}$, $F_{collision}$, and $F_{view}$, respectively; for finding out-of-road violations, the three NNs are trained to predict $F_{wronglane}$, $F_{offroad}$, and $F_{deviation}$, resp. The NNs all have one hidden layer with size 100. The batch-size, training epoch and optimizer are set to 200, 200, and the Adam optimizer.} 

\newedit{AV-FUZZER \cite{avfuzzer} first runs a global \GA for several iterations and enters a local \GA with the initial population set to the scenario vectors with the highest fitness scores. It also starts a new global \GA every time when the fitness score of the current generation does not increase anymore compared with a running average of the last five generations. We keep the hyper-parameters used as in the original implementation e.g. population size is set to 4.}

\newedit{We did not directly compare with FITEST \cite{interactiontest}, Asfault \cite{Kuutti2020} or FusionFuzz\cite{zhong2021detecting} since they are essentially \GA with specifically designed objectives targeting testing of the integration component of an \av, a controller's performance under different road networks, or the fusion component of an \av, respectively, while we focus on testing a black-box end-to-end system on a predefined map available with different specific scenarios by mutating different elements (\eg weather, agents, their positions and behaviors)}.

%% file: body/7_results.tex
\section{Results}
\label{sec:results}

To evaluate how efficiently \tool can find unique \bugs, we explore the following research questions:

\textbf{RQ1: Evaluating Performance.} How effectively can \tool find unique \sout{\bugs in comparison to}\new{violations versus} baselines?

\textbf{RQ2: Evaluating Design Choices.} What are the impacts of different design choices on \tool?

\textbf{RQ3: Evaluating Repair Impact.} Can we leverage \bugs found by \tool to improve the controller?

\newedit{\textbf{RQ4: Evaluating Generalizability.} Can \tool generalizes to a different system and simulator combination?}

\input{rqs/rq1}
\input{rqs/rq2}
\input{rqs/rq3}
\input{rqs/rq4}

%% file: rqs/rq1.tex
\begin{figure}[t]
{\centering
    
    {\includegraphics[width=0.15\textwidth]{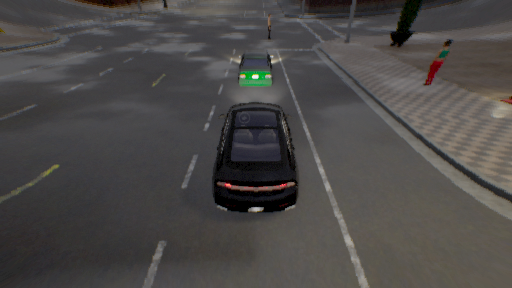}}
    {\includegraphics[width=0.15\textwidth]{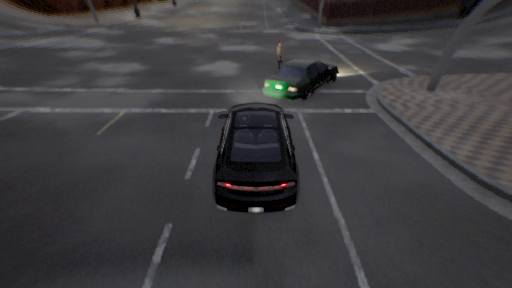}}
    {\includegraphics[width=0.15\textwidth]{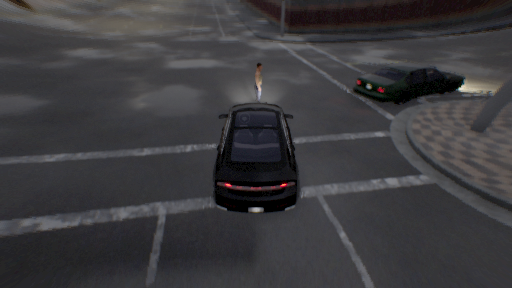}}
    
    \vspace{2mm}
    {\includegraphics[width=0.15\textwidth]{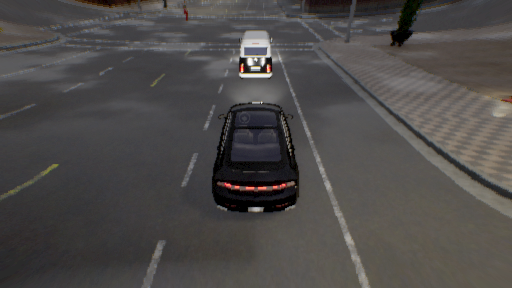}}
    {\includegraphics[width=0.15\textwidth]{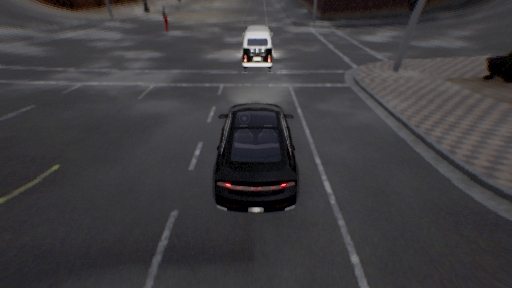}}
    {\includegraphics[width=0.15\textwidth]{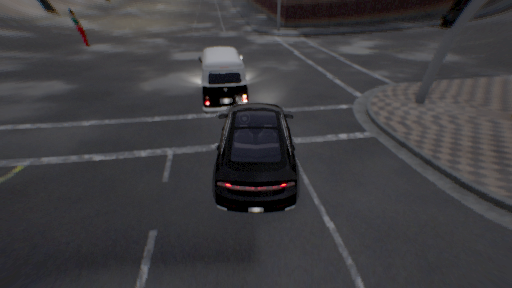}}
    
    \vspace{2mm}
    {\includegraphics[width=0.15\textwidth]{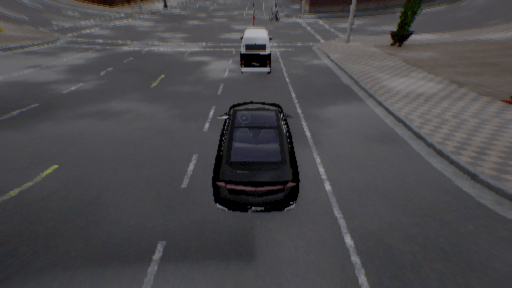}}
    {\includegraphics[width=0.15\textwidth]{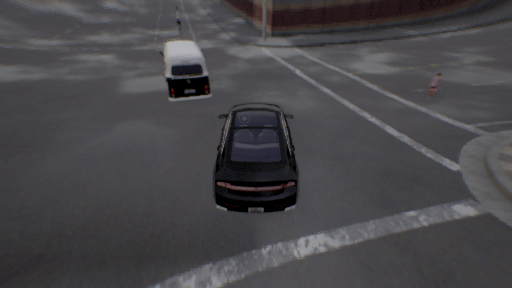}}
    {\includegraphics[width=0.15\textwidth]{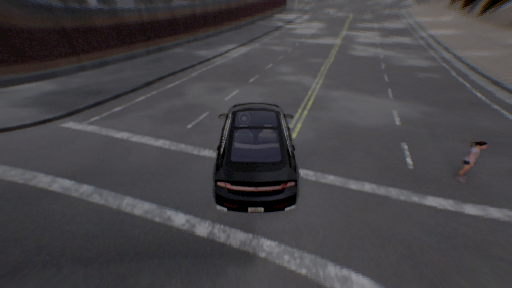}}

\vspace{-2mm}
\caption{\textbf{\small{RQ1.~Example \bugs found by \tool.}}}
\label{fig:bugs_demo}
}
\begin{FlushLeft}
\textbf{{\small For each row, the time goes by from left to right. (1st row) pid-1 controller collides with a pedestrian crossing the road. (2nd row) pid-2 controller collides with the stopped leading car. (3rd row) lbc controller makes a wide turn into the opposing lane (considered "off-road").}}
\end{FlushLeft}

\end{figure}
\begin{figure}[t]
\centering
\includegraphics[width=0.48\textwidth]{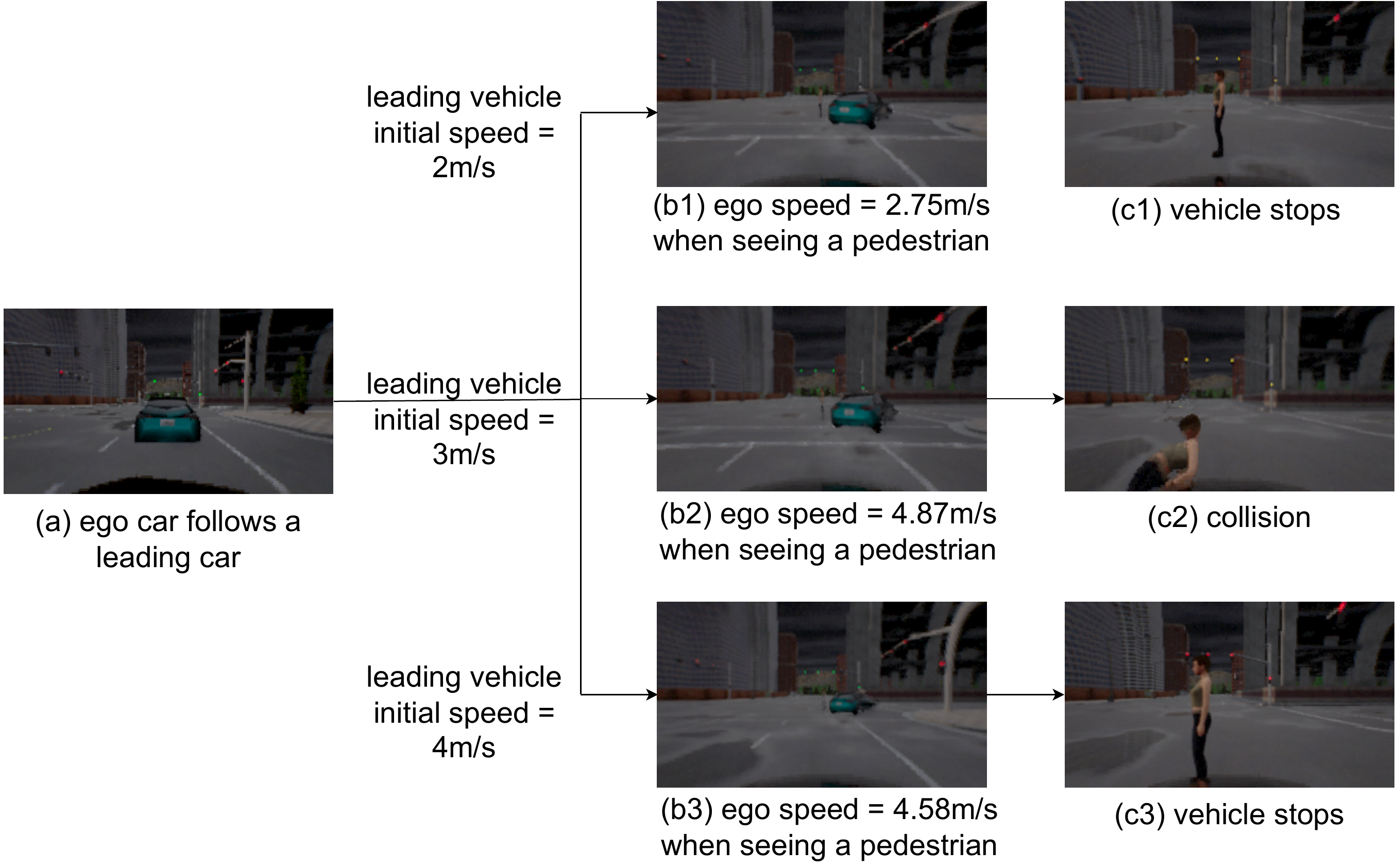}
\caption{\small{\textbf{\new{An example (front camera view) where a small parameter change leads to distinct outcomes for lbc in \carla.}}}}
\label{fig:variation_demo}
\end{figure}

\smallskip
\RQ{1}{Evaluating Performance.}
We first explore whether \tool can find realistic and unique \bugs for the \av controllers under test. 
Note that all the \bugs are generated by valid \specifics, as they are created using \carla's API interface (\newedit{we also randomly spot-checked 1000 of them}). 
We run \tool with \GAUNNNGRAD on all three controllers for 700 simulations, with the search objective to find collision \bugs in the town05 \ls. Note that even though the search objective is set to finding collisions, the process might also find a few off-road \bugs. Overall, \tool found 725 unique \bugs total across the three controllers for this \ls. In particular, it found 575  unique \bugs for the \textbf{lbc} controller, 80 for the \textbf{pid-1} controller, and 70 for the \textbf{pid-2} controller. Since \textbf{pid-1} and \textbf{pid-2} assume extra knowledge of the states of other environment objects, it is usually harder to find \bugs.  
Figure~\ref{fig:bugs_demo} shows snapshots of example \bugs found by \tool.
These examples illustrate that starting from the same \ls, different violations can be generated because of the high-dimensional input feature space. 
ks stucgetFigure~\ref{fig:variation_demo} shows an example \newnew{violation of the motivating \ls "Turning right while leading car slows down/stops".} \remove{where a}\newnew{A} small mutation of the leading vehicle's speed from 3m/s to 4m/s\newnew{ or 2m/s} leads to completely different outcomes. When the leading vehicle's speed is 3m/s, the ego car has less time to detect the presence of a pedestrian obstructed by the leading vehicle \newnew{(b2)}\remove{(b1)} and ends up with colliding with the pedestrian \newnew{(c2)}\remove{(c1)}. When the leading vehicle's speed is 4m/s, the ego car detects the pedestrian earlier \newnew{(b3)}\remove{(b2)} and avoids the collision by braking on time \newnew{(c3)}\remove{(c2)}. \newnew{When the leading vehicle's speed is 2m/s, although the ego car detects the pedestrian late (b1), the ego car is at low speed (2.75m/s) so it also manages to avoid the collision (c1).} It should also be noted that the collision \newnew{(c2)}\remove{(c1)} can be avoided if the ego car brakes the first time it sees the pedestrian \newnew{(b2)}\remove{(b1)}. \newnew{This \bug is non-trivial to be found since the change of the leading NPC vehicle's initial speed leads to different reaction of the ego car and it is not clear what value of the initial speed along with other parameters in the search space leads to the collision. In fact, \AVFUZZER fails to find this violation since \AVFUZZER gets stuck at another \bug involving the ego car's collision with the slowing down leading NPC vehicle.}



\begin{figure}[t]
\centering
\includegraphics[width=0.24\textwidth]{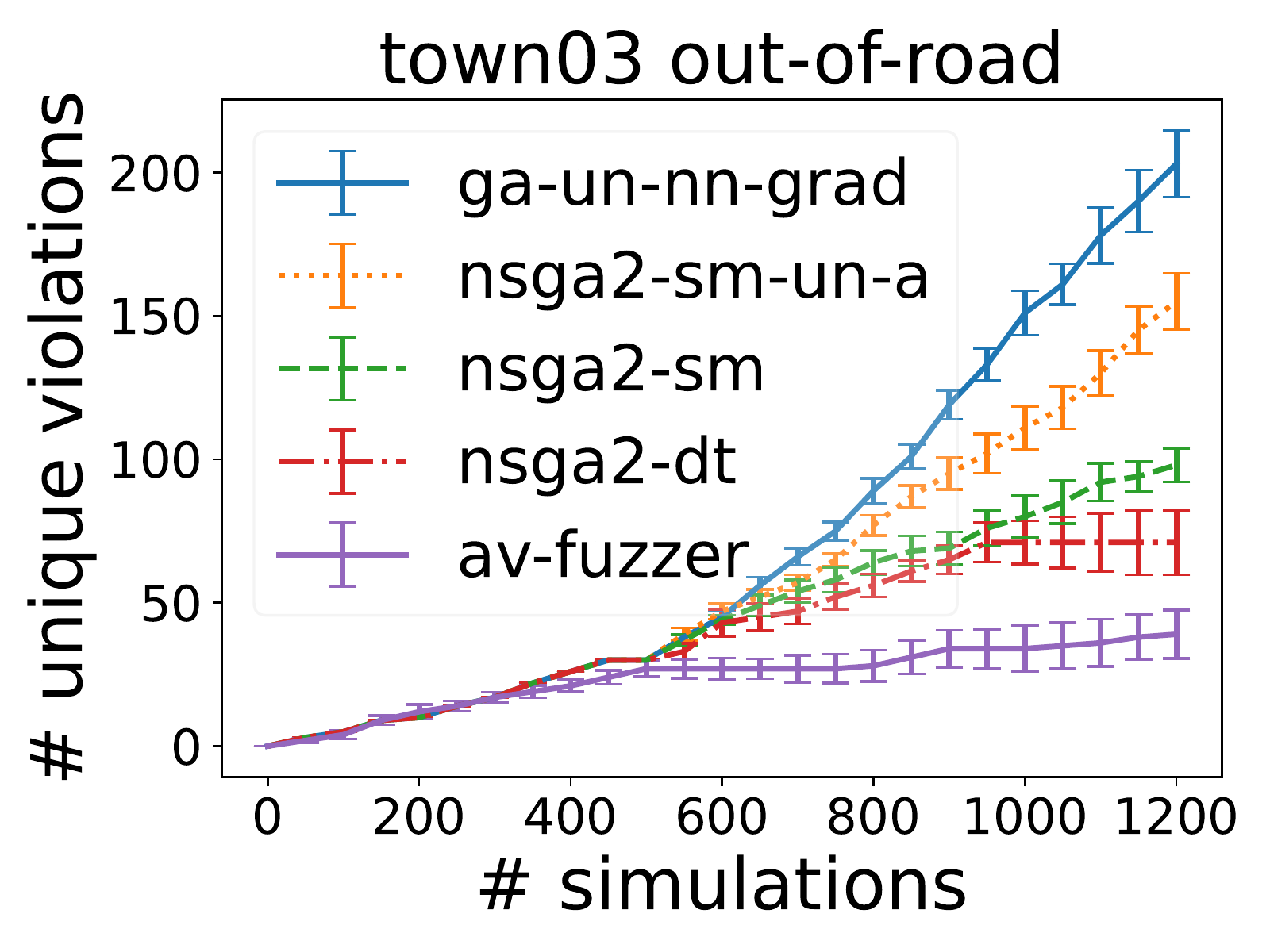}
\includegraphics[width=0.24\textwidth]{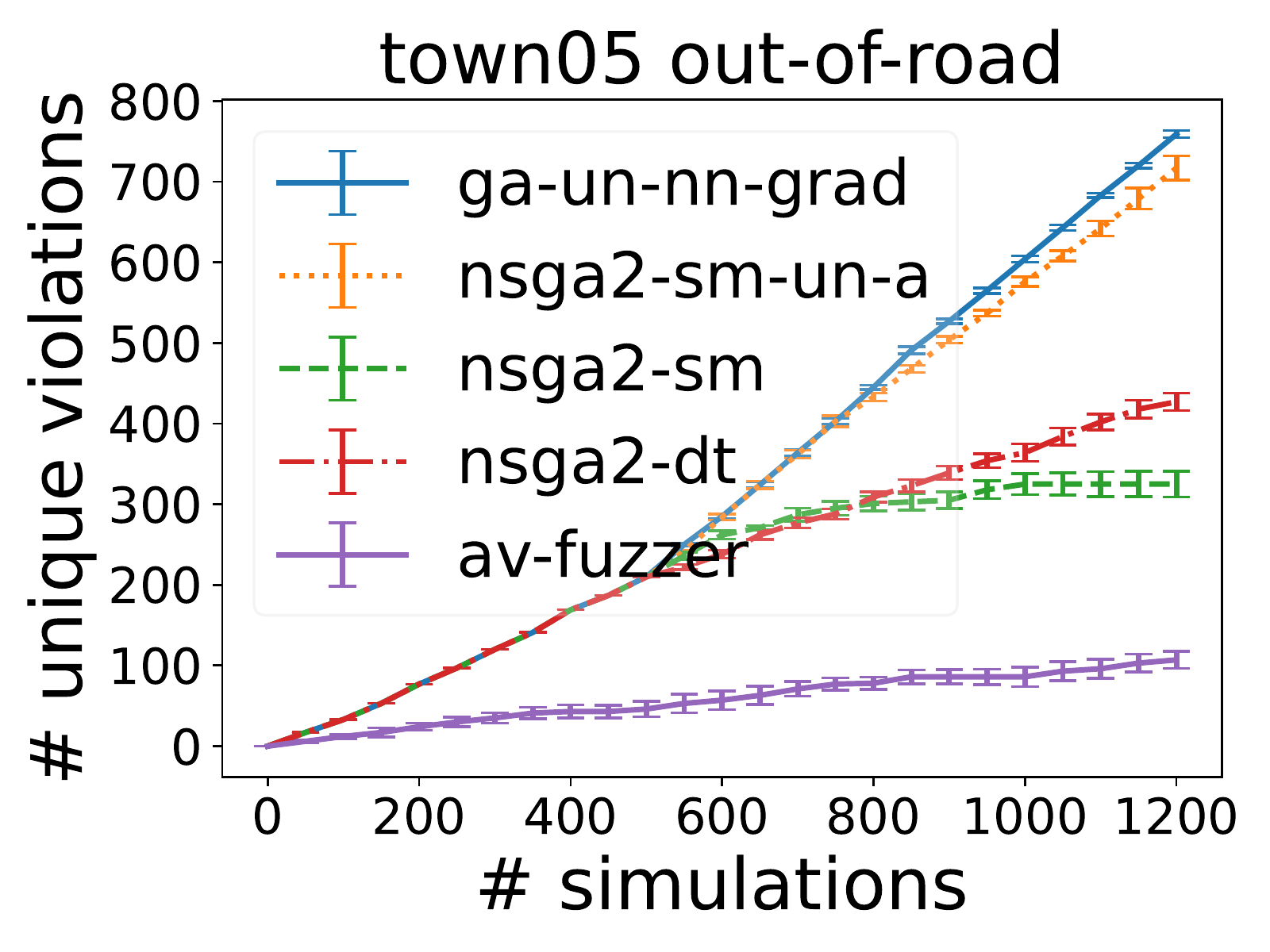}
\includegraphics[width=0.24\textwidth]{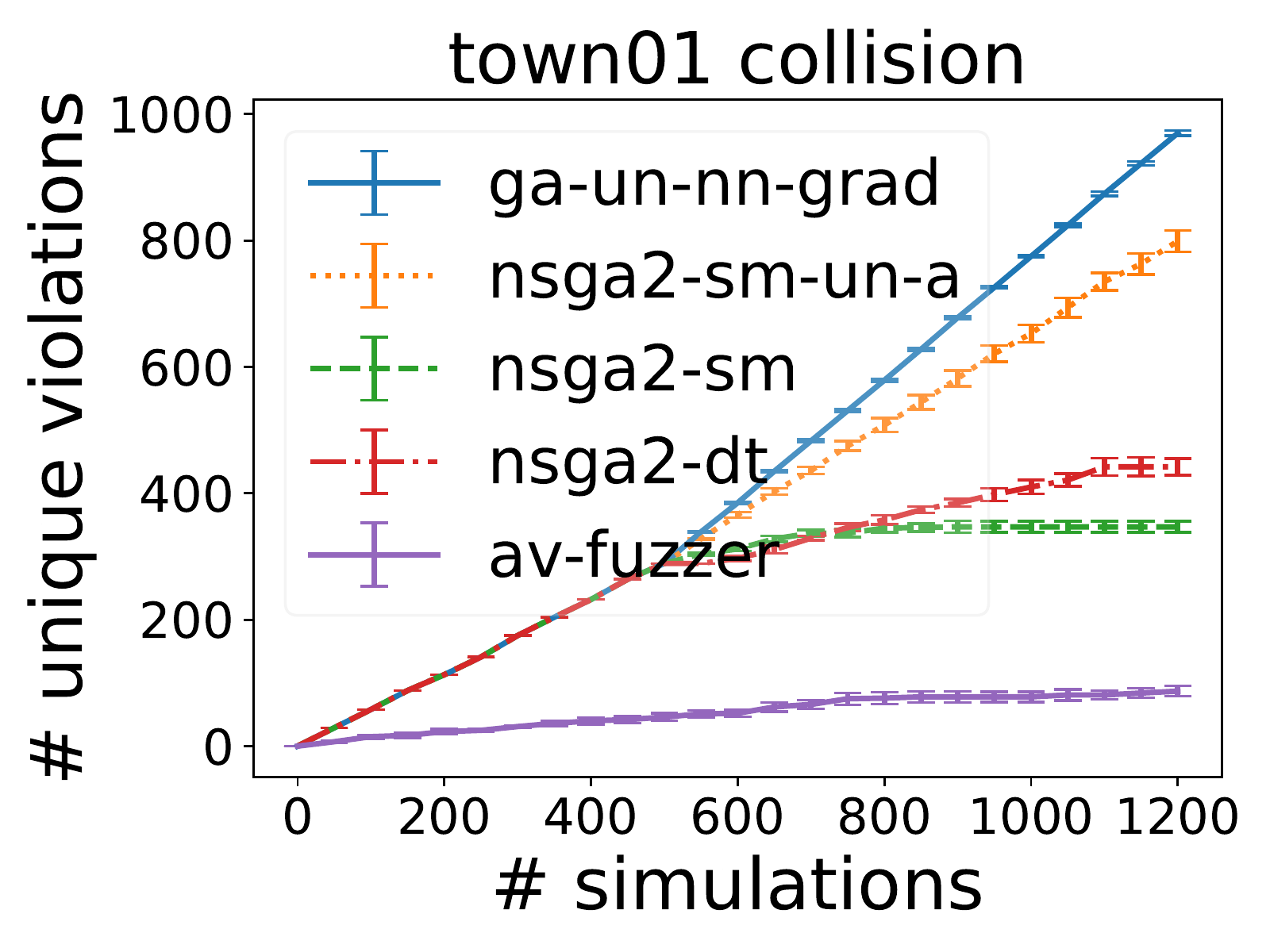}
\includegraphics[width=0.24\textwidth]{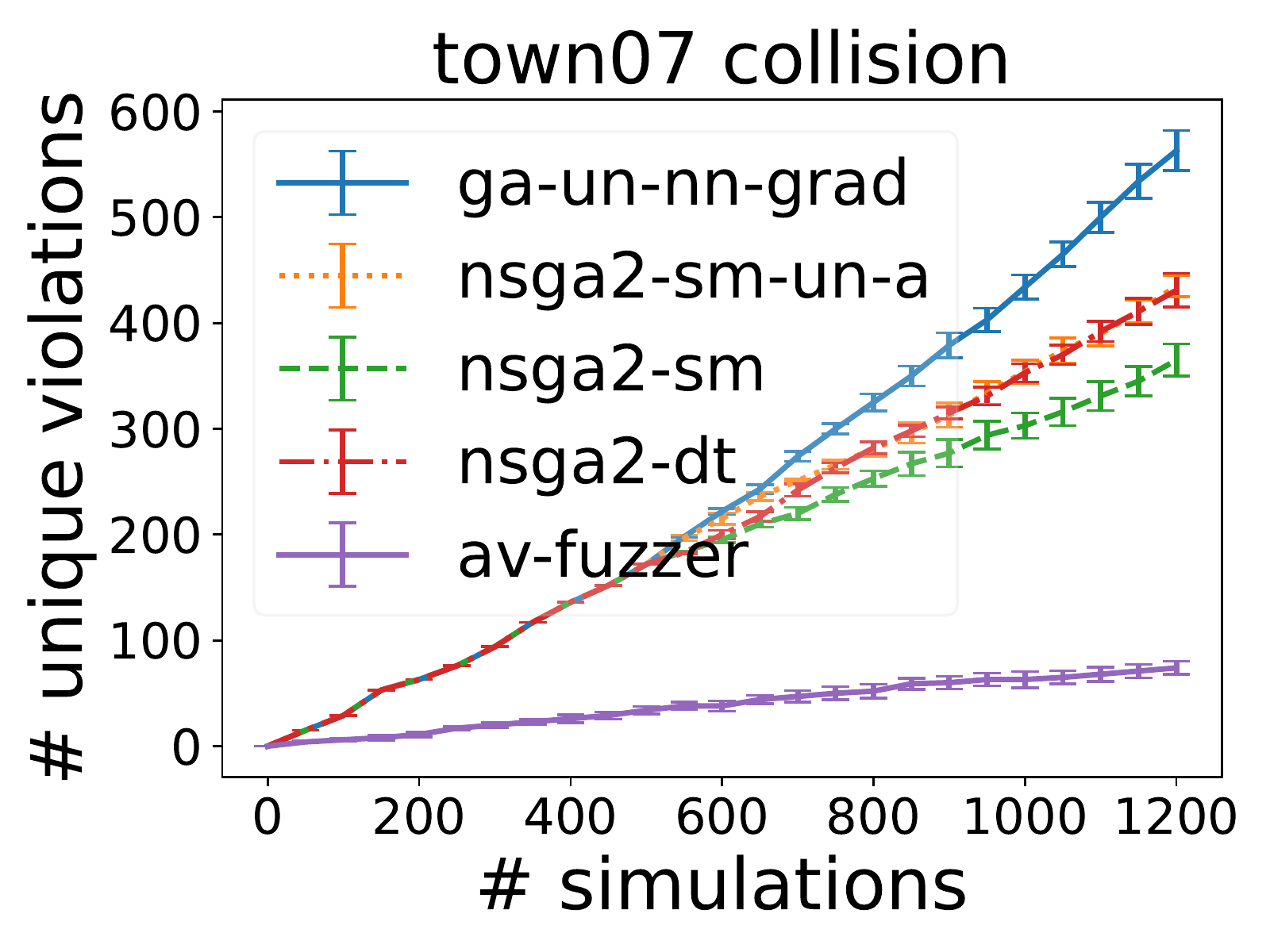}

\caption{\small{\textbf{RQ1.~average \# unique off-road or collision violations.}}}
\label{fig:search_methods_on_different_scenes}
\end{figure}


We compare \tool (\ie \GAUNNNGRAD) with the baseline methods \NSGADT, \NSGASM, \remove{and }\NSGAUNSMA\newnew{, and \AVFUZZER} under four different \lss. We focus on collision \bugs for two logical scenarios and off-road \bugs for the other two. In each setting, we run each method 6 times and report mean and standard deviation. \newnew{For \AVFUZZER, we fuzz for 1200 simulations. For other methods, we}\remove{We also} assume 500 pre-collected seeds and fuzz for 700 simulations.  Figure~\ref{fig:search_methods_on_different_scenes} shows the results.

\GAUNNNGRAD consistently finds 10\%-39\% more than the \newnew{best-performing }baseline method\remove{s}. In particular, \GAUNNNGRAD finds 41, 51, 135 and 111 more unique \bugs over the second-best method in the four \edit{\lss}. 

\newedit{We further conduct Wilcoxon rank-sum test \cite{Wilcoxon} and Vargha-Delaney effect size test \cite{VDtest, guidetostatstest}. For all the settings, the 90\% confidence interval of the effect size between \GAUNNNGRAD and the best baseline is (0.834, 1.166) meaning large effect size, and the p-value is $3.95e^{-3}$ suggesting the gain of the proposed method is statistically significant.}

After collecting all the violation-producing \edit{\specificss}, we measure how many are truly unique as per our uniqueness criteria. \GAUNNNGRAD and \NSGAUNSMA win by a large margin. For example, for the turning left non-signalized junction \edit{\ls}, \GAUNNNGRAD and \NSGAUNSMA have $100\%$ unique violations while the other \newnew{three}\remove{two} methods have only $42\%$\newnew{,}\remove{and} $22\%$\newnew{ and $10$\%}. This is expected since they both have a duplicate elimination component inherent to the search strategy. The results show that the baselines NSGA2-SM\newnew{,}\remove{and} NSGA2-DT\newnew{, and \AVFUZZER} waste many resources by running similar violation-producing \edit{\specificss}.

After introducing duplicate elimination (UN) and incremental learning (A), \NSGAUNSMA finds more violations than \NSGASM. But \GAUNNNGRAD still has advantages: \sout{1.}\new{(i)} Our goal is to maximize the number of unique \bugs than finding \bugs with the best Pareto front \cite{testing_vision18, nsga2nn}, so a binary classification NN gives a better guide than multiple regression NNs. \sout{2.}\new{(ii)} The constrained gradient-guided permutation gives a further boost. The second point is\sout{ also} shown in the ablation study in RQ \ref{sec:rq_design_choice}. \newnew{Besides, we have observed that \AVFUZZER finds much fewer \bugs. It even finds fewer unique \bugs than the seed collection stage (for which \GAUN is used) of other methods. The main reason is that \AVFUZZER has very limited diversity exploration. In particular, its default mutation rate is small and its local \GA starts with the mutated duplicates of the global best scenario vector so far, both of which limit diversity. If the global best scenario vector does not change after several generations, all the local \GA will start with the same duplicates. Moreover, its resampling process picks the farthest scenario vectors from the existing ones but does not consider the distances among the selected scenario vectors, which results in restarting at a local cluster of scenario vectors with limited diversity.}

\begin{figure}[ht]
\centering
    {\includegraphics[width=0.15\textwidth]{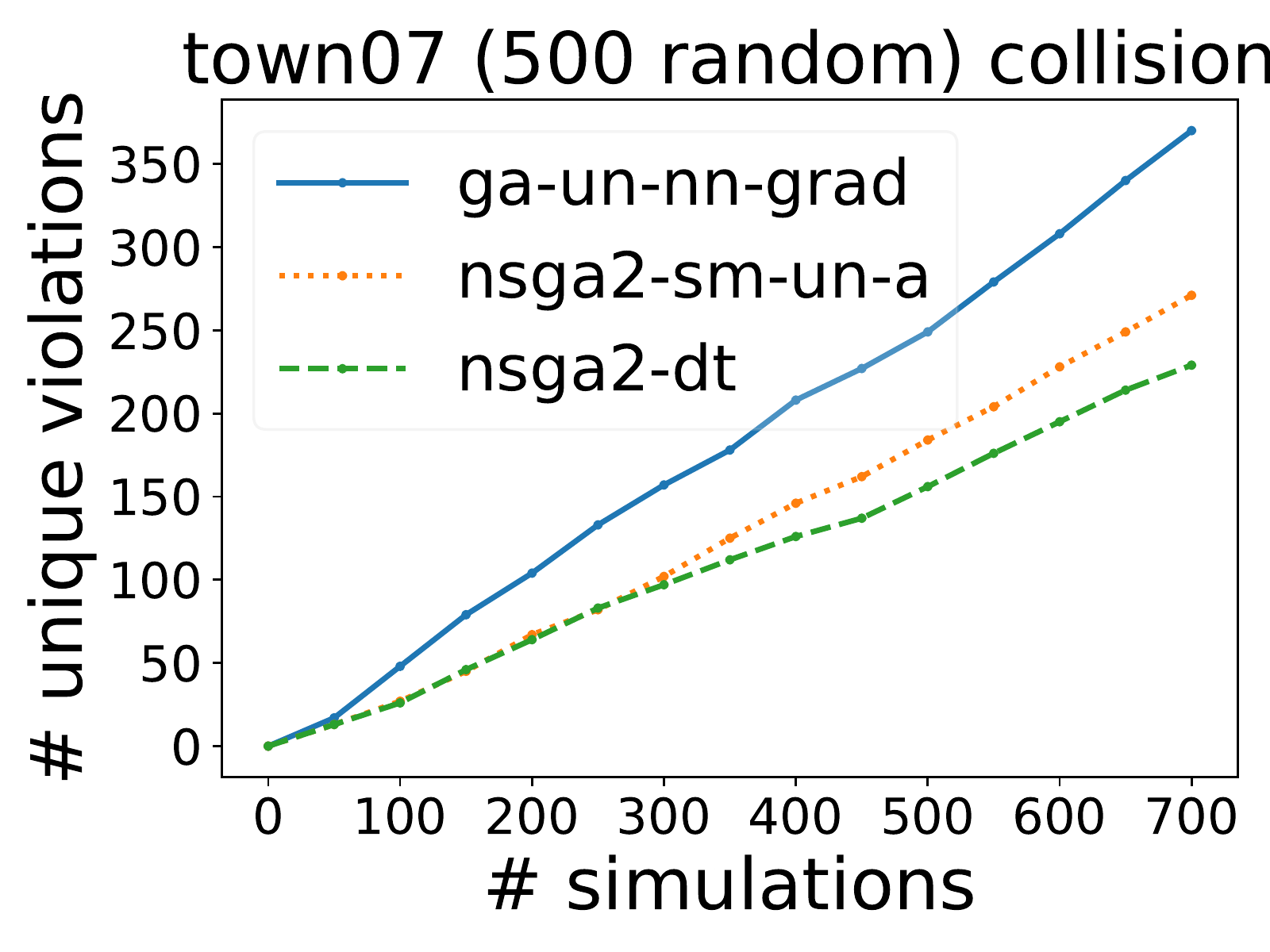}}
    {\includegraphics[width=0.15\textwidth]{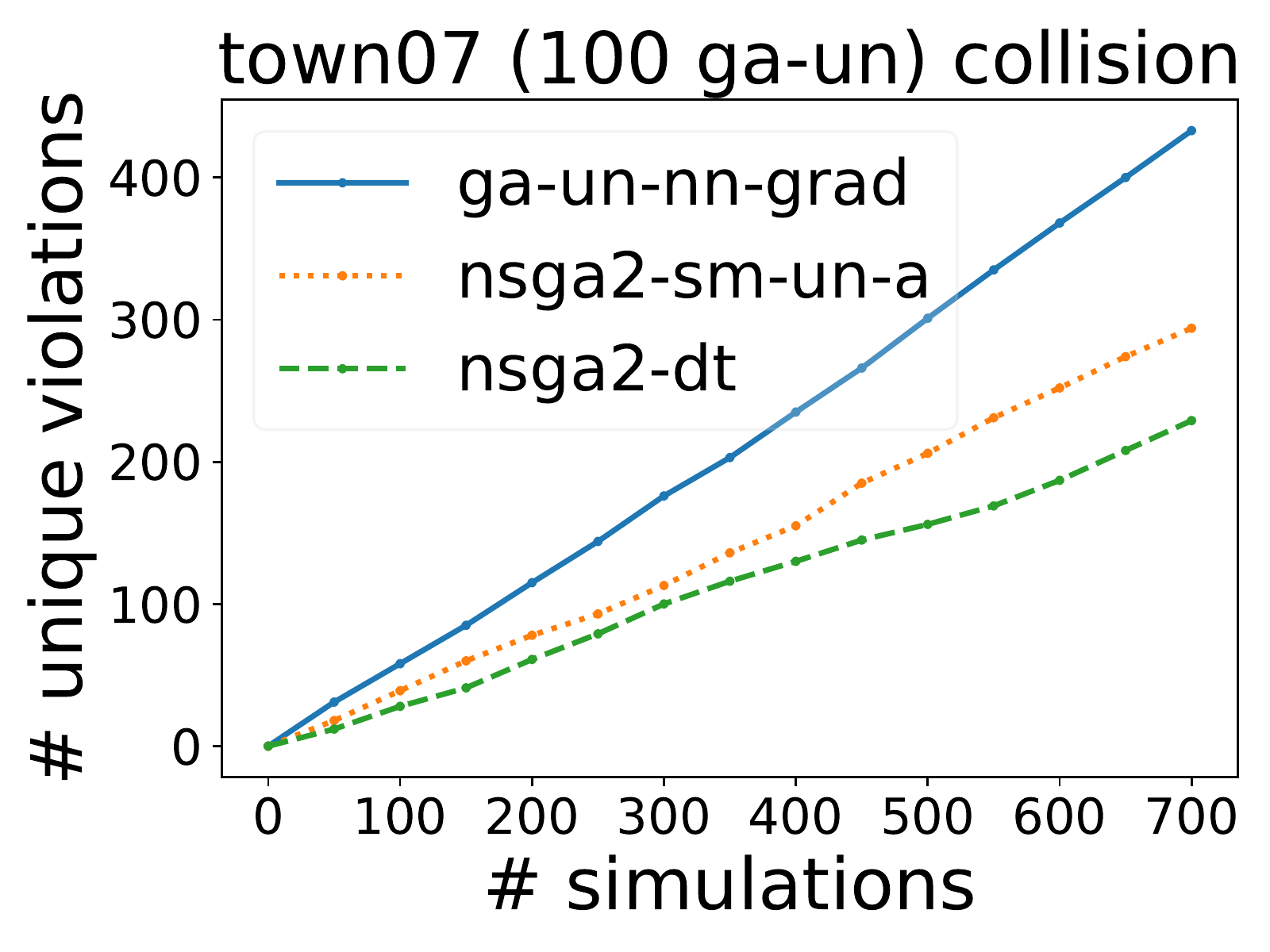}}
    {\includegraphics[width=0.15\textwidth]{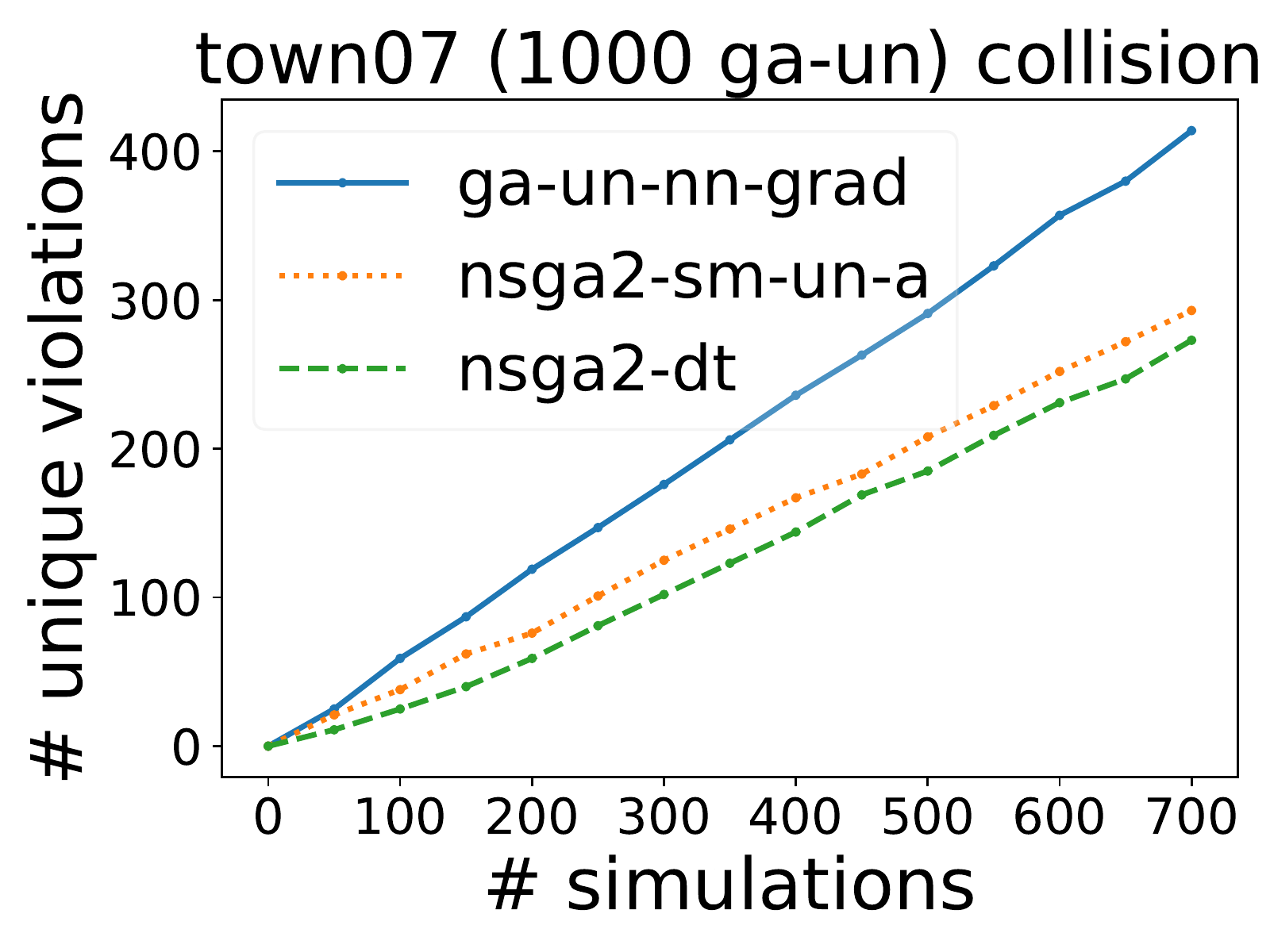}}
    \vspace{-2mm}
\caption{\small{\textbf{RQ1.~\# unique violations under different initial seeds.
}}} 
\label{fig:ablation_initial_seed_town07_collision}
\vspace{-1mm}
\end{figure}

Next, we study if \GAUNNNGRAD can effectively find more unique \bugs over baselines under different initial seeds. We compare the number of unique \bugs found by \GAUNNNGRAD with \NSGAUNSMA and \NSGADT for 700 simulations, assuming 500 initial seeds collected by \RA, and 100 and 1000 initial seeds collected by \GAUN, resp. As shown in \Cref{fig:ablation_initial_seed_town07_collision}, \GAUNNNGRAD finds 99, 139, and 121 more unique \bugs than the baselines.

\RS{1}{\tool finds hundreds of unique \bugs across all three controllers. 
On average, it finds 9\%-41\% more unique violations over the second-best baseline. }


%% file: rqs/rq2.tex

\smallskip
\RQ{2}{Evaluating Design Choices.}
\label{sec:rq_design_choice}
We study the influence of each component and choice of hyper-parameters on \tool. 
We present the results \new{for}\sout{with} the town07 \edit{\ls}, with finding collisions as the\sout{ search} objective. 
However, the observations also hold in general for other \edit{\lss} and objectives.  


\begin{figure}[ht]
\centering
    {\includegraphics[width=0.24\textwidth]{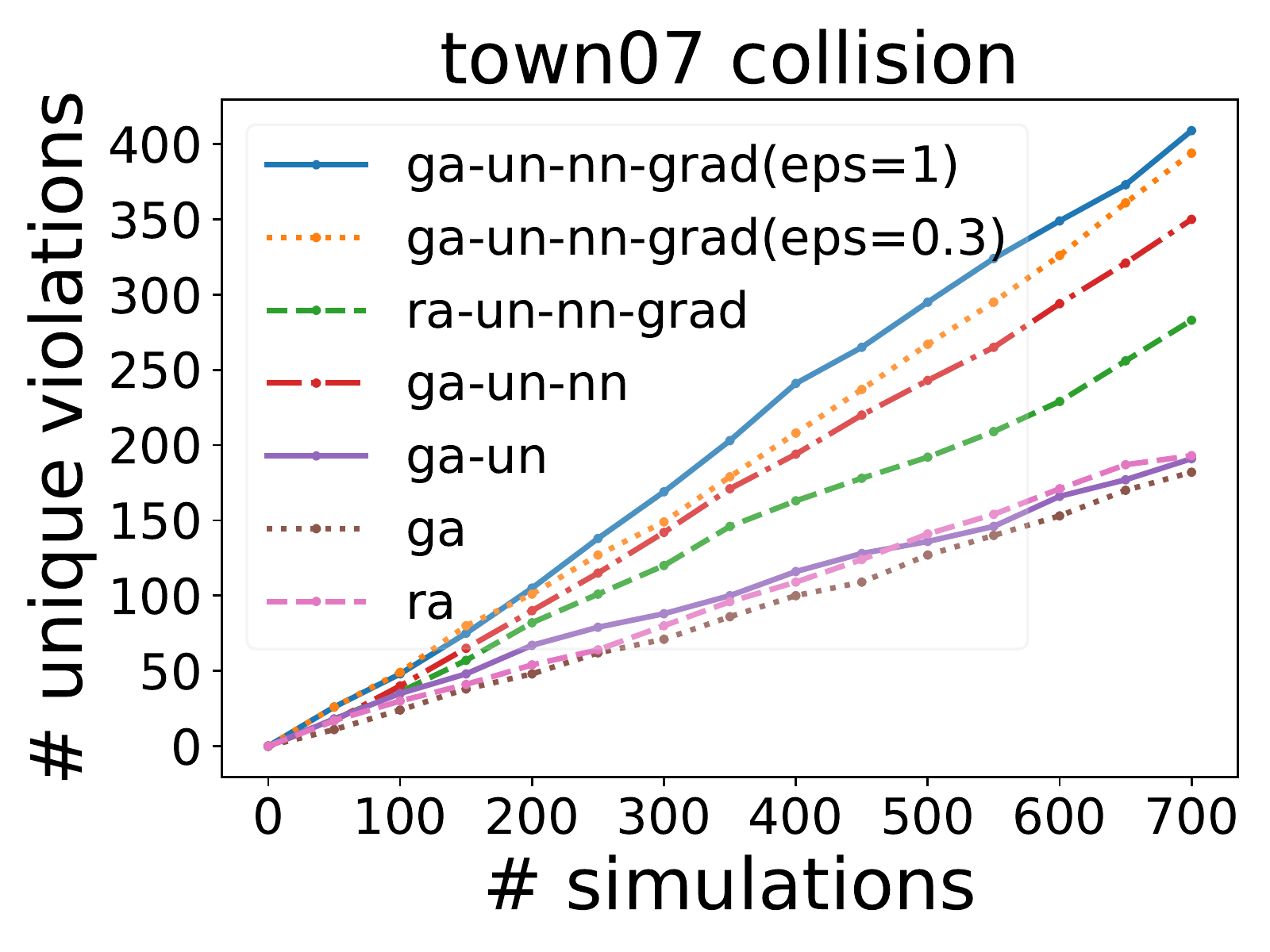}}
\caption{\small{\textbf{\#unique \bugs found by \tool's variants. 
}}}
\label{fig:ablation_town07_collision}
\end{figure} 

We conduct an ablation study on the impact of each \GAUNNNGRAD component, comparing the number of unique \bugs found by \GAUNNNGRAD with the six variations shown in Table~\ref{tab:all_methods}. 
\Cref{fig:ablation_town07_collision} presents the results. 


\noindent
\textit{- \GAUNNNGRAD ($\epsilon = 1$ vs. $0.3$).} With larger $\epsilon$, slightly more violations are detected. A larger $\epsilon$ value can perturb the input with a larger magnitude. 
Thus, it can have more diverse seeds and reach a better optimum in terms of violations likelihood considered by the  NN used for seed-selection and mutation.  

\noindent
\textit{- \GAUNNNGRAD vs. \RAUNNNGRAD.} \GAUNNNGRAD finds more violations indicating the importance of the base sampling strategy. 

\noindent
\textit{- \GAUNNNGRAD vs. \GAUNNN vs. \GAUN.} \GAUNNNGRAD finds more unique violations than \GAUNNN and \GAUNNN beats \GAUN. These show the necessity of the gradient-guided mutation component (GRAD) and seed selection component (NN). 
Furthermore, \GAUN finds slightly more unique \bugs than \GA. 

We next explore the sensitivity of different search methods under nine different combinations of uniqueness thresholds, $th_1$ and $th_2$, as discussed in~\Cref{sec:step3}. We compare them for 300 simulations after the initial seed collection stage. The trend also holds for more simulations. Table \ref{tab:diff_th} shows \GAUNNNGRAD finds at least 10-30\% more unique \bugs than the second-best baseline method under seven settings. For the setting (10, 75) and (20, 75), none of the methods can find new \bugs. This is because the uniqueness constraint is too stringent, so the sampling component cannot find a valid sample that obeys the constraint.



\begin{table}[ht]
    \footnotesize
    \centering
        \caption{\textbf{\small{\sout{Number}\new{\#} of unique \sout{\bugs}\new{violations} found \sout{by each search method }under different \sout{definitions of unique \bugs}\new{$th_2,th_1$}.}}}
    \label{tab:diff_th}
    \vspace{-2mm}
    \begin{tabular}{l|l|l|l}
    \toprule
        ($th_2$,$th_1$) & \GAUNNNGRAD & \NSGAUNSMA & \NSGADT \\
    \toprule
        (5, 25) & \underline{175} & 110 & 138  \\
        (10, 25) & \underline{168} & 121 & 142  \\
        (20, 25) & \underline{161} & 109 & 131  \\
        \midrule
        (5, 50) & \underline{173} & 121 & 146  \\
        (10, 50) & \underline{169} & 131 & 92  \\
        (20, 50) & \underline{35} & 31 & 16  \\
        \midrule
        (5, 75) & \underline{26} & 16 & 1 \\
        (10, 75) & 0 & 0 & 0 \\
        (20, 75) & 0 & 0 & 0 \\
    \bottomrule
    \end{tabular}
\end{table}

\RS{2}{Each component of \GAUNNNGRAD contributes to the final superior performance and combined they find more unique \bugs compared to all other settings. }

%% file: rqs/rq3.tex
\smallskip
\RQ{3}{Evaluating Impact on Repair.}
Since the purpose of finding erroneous behavior in any software is to help with removing the errors\sout{. We}\new{, we} speculated whether we can leverage the \bugs found to improve a controller to reduce future \bugs. We focus on the collisions found for four \edit{\lss}. For each one, we randomly select 200 detected \bugs by \GAUNNNGRAD for \textbf{lbc}, and split the corresponding \edit{\specificss} into 100 for retraining and 100 for testing. We use  \textbf{pid-1} as a teacher model to run the 100 \edit{\specificss} for retraining and collect the camera data where it finishes successfully. The collected camera images are down-sampled to two frames per sec (about 2000 images) and use them to fine-tune the \textbf{lbc} model for one epoch.  Finally, we test the retrained model on the held-out 100  \edit{previously failing \specificss}. \Cref{tab:fixing} shows that the retrained controller succeeds in \sout{driving through for }over 75\% of the originally failing \edit{\specificss}.


{
\setlength{\tabcolsep}{3pt}
\renewcommand{\arraystretch}{1}
\vspace{0.2cm}
    
\begin{table}[h]
\centering
\footnotesize
\caption{\textbf{\small{\sout{Number}\new{\#} of\sout{ traffic} violations fixed in the held-out dataset.}}}
 \label{tab:fixing}

    \begin{tabular}{l|r|r}
    \toprule
          \edit{\lss names} & \# retraining  & \# violations  \\
         &     data       &  fixed \\
    \toprule
         turning right while leading car slows down & 64 & 82 / 100 \\
         turning left non-signalized & 47 & 76 / 100 \\
         crossing non-signalized & 91 & 100 / 100 \\
         changing lane & 64 & 75 / 100 \\
    \bottomrule
    \end{tabular}
    
\end{table}
}

\RS{3}{In our preliminary study, retraining with \bugs found by \tool improved the \textbf{lbc} controller's performance on failure cases by 75\% to 100\%.\vspace{-2mm}}

%% file: rqs/rq4.tex
\smallskip
\RQ{4}{Evaluating Generalizability.}

\begin{figure}[ht]
\centering
    {\includegraphics[width=0.24\textwidth]{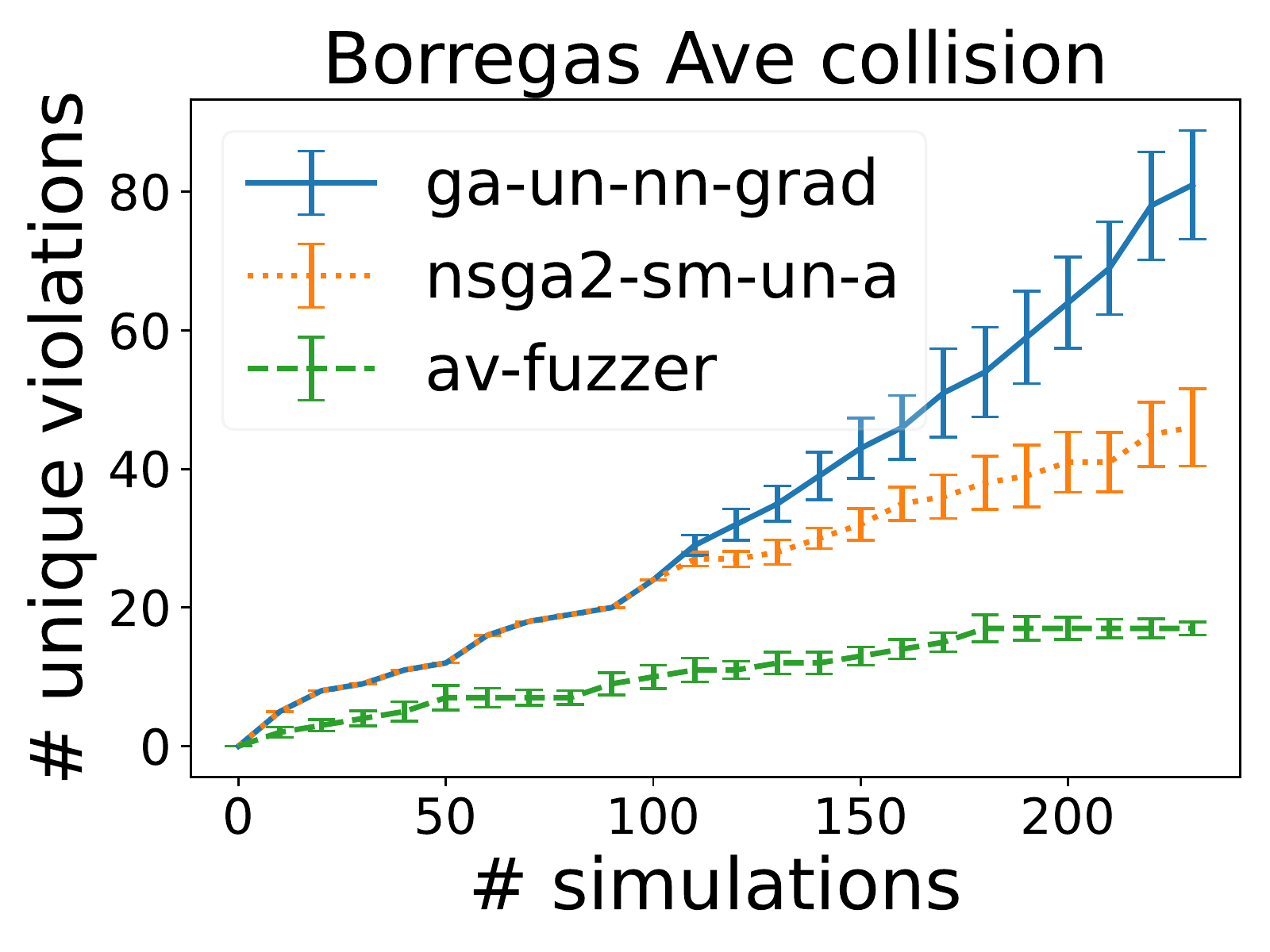}}
\caption{\small{\textbf{
\newedit{RQ4. average \#unique collision \bugs.}
}}}
\label{fig:apollo_num_collision_found}
\end{figure}

\newedit{In \Cref{sec:results} we reported experimental results based on a single simulator, \carla, and three research-oriented controllers. To evaluate the generalizability of \tool, we conduct a preliminary study on \apollo, an industrial-grade \av{} controller \cite{apollo}, using a different simulator, \svl (version 2021.3) \cite{svl,svl-paper}. We analyze the \svl API similarly to \carla (\Cref{sec:grammar}) and focus on collision \bugs (\Cref{subsec:obj}).
We use a logical scenario where the ego car conducts a left turn at a signalized junction while another vehicle comes from the other side and a pedestrian crosses the street. Since the search space has 11 parameters (we do not consider parameters like weather and lighting since their implementations in \svl do not influence LiDAR which \apollo mostly relies on for its perception module) to search for, to speed up the convergence of the search process, we reduce the population size to 10. All other hyper-parameters and settings are kept the same as in RQ1. We run \tool and the best performing baseline \NSGAUNSMA for 14 generations totaling 140 simulations (excluding an initial 100 \sout{warm-up}simulations \new{for the seed collection stage}) and run \AVFUZZER for 240 simulations (since it does not have \sout{warm-up simulations}\new{a seed collection stage}). We then compare them over the entire 240 simulations. As shown in \Cref{fig:apollo_num_collision_found}, on average of six repetitions, \GAUNNNGRAD finds 76 unique \bugs --- which is 49\% and 375\% more, respectively, than the two baseline methods \NSGAUNSMA and \AVFUZZER (51 and 16 unique \bugs, resp.). 
We further conduct Wilcoxon rank-sum test and Vargha-Delaney effect size test. The 90\% confidence interval of the effect size between \GAUNNNGRAD and the best baseline is (0.807, 1.165) meaning large effect size, and the p-value is $5.07e^{-3}$ suggesting the gain of the proposed method is statistically significant.}

\begin{figure}[th]
\centering
    {\includegraphics[width=0.15\textwidth]{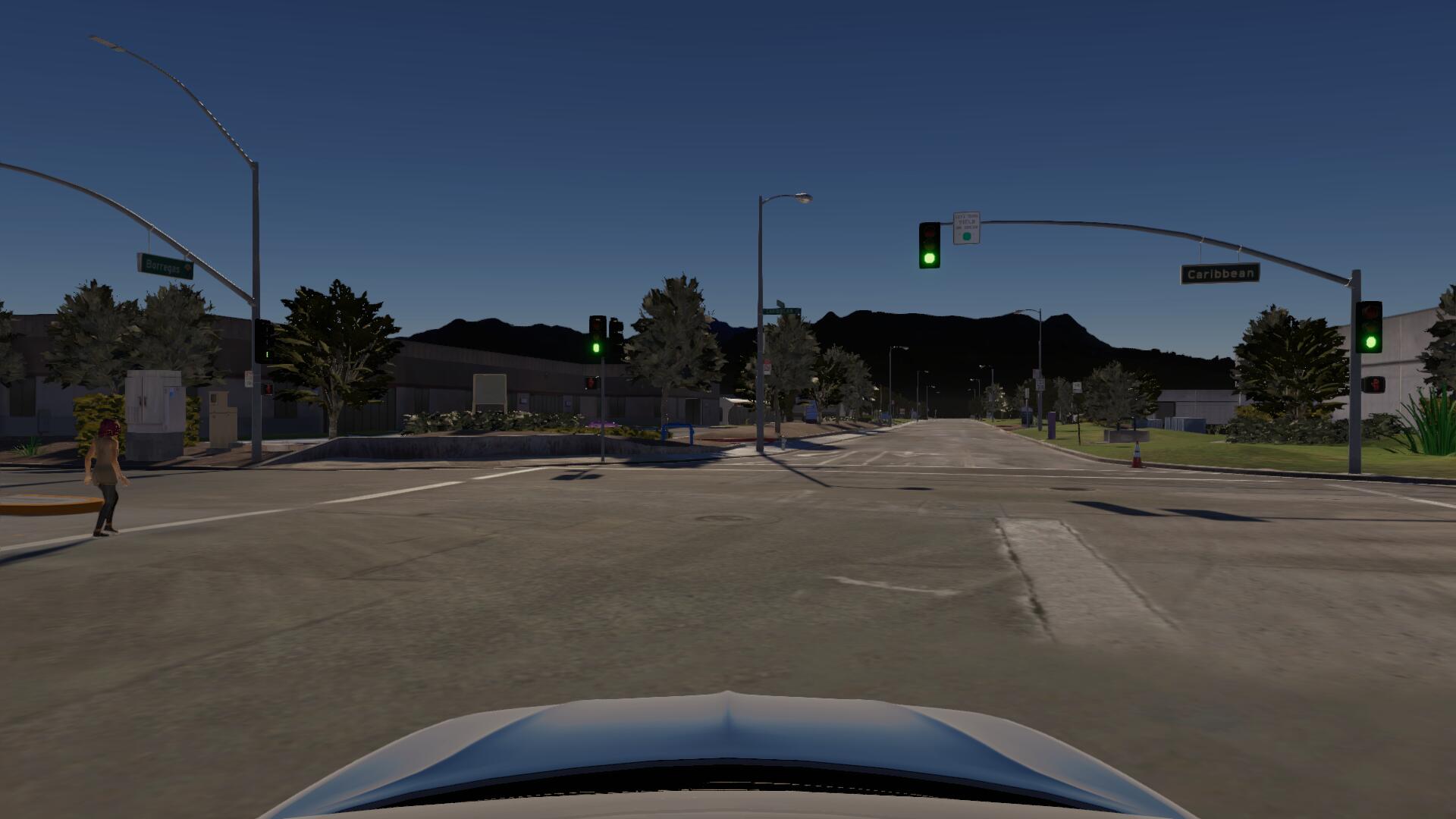}}
    {\includegraphics[width=0.15\textwidth]{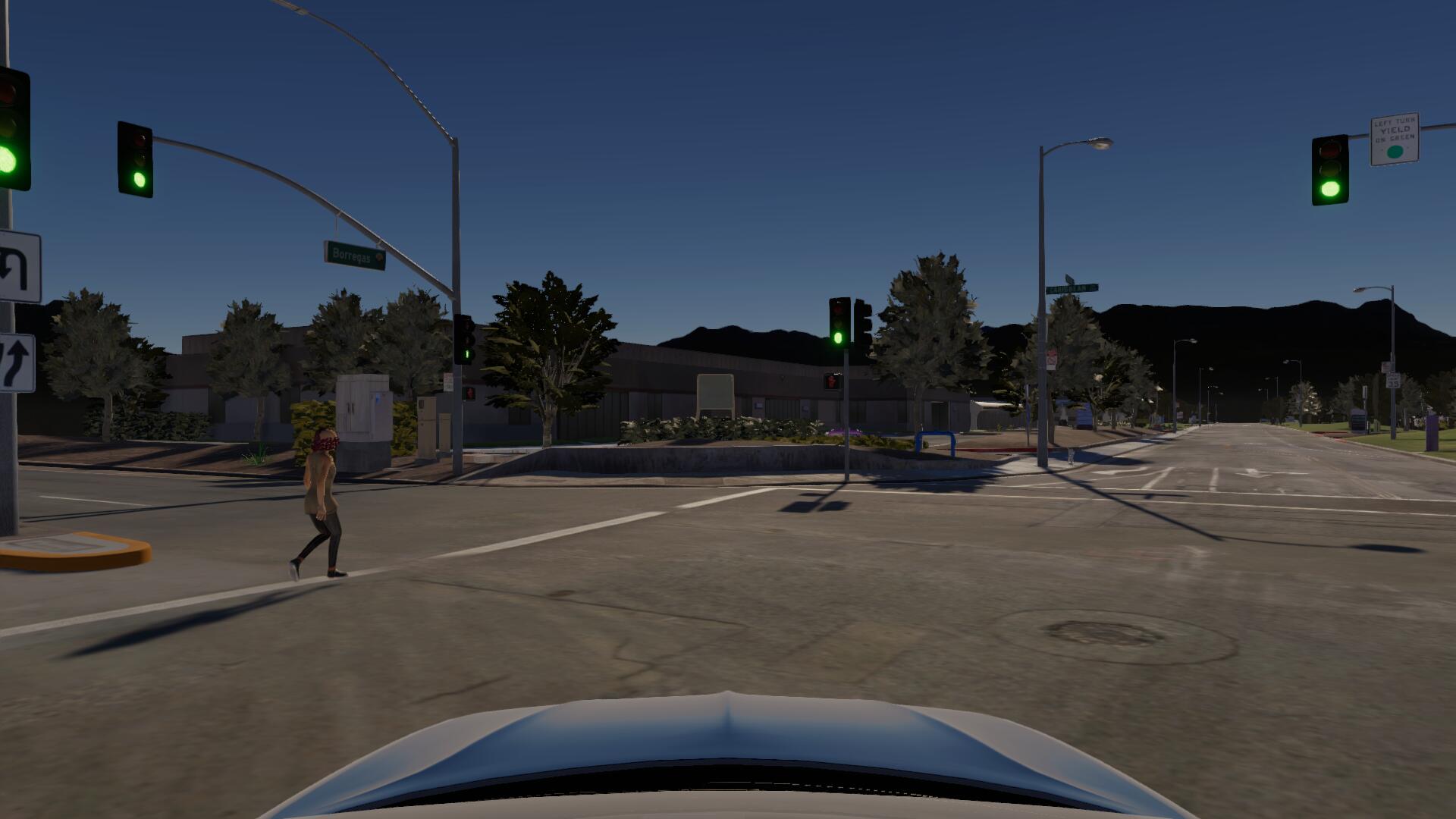}}
    {\includegraphics[width=0.15\textwidth]{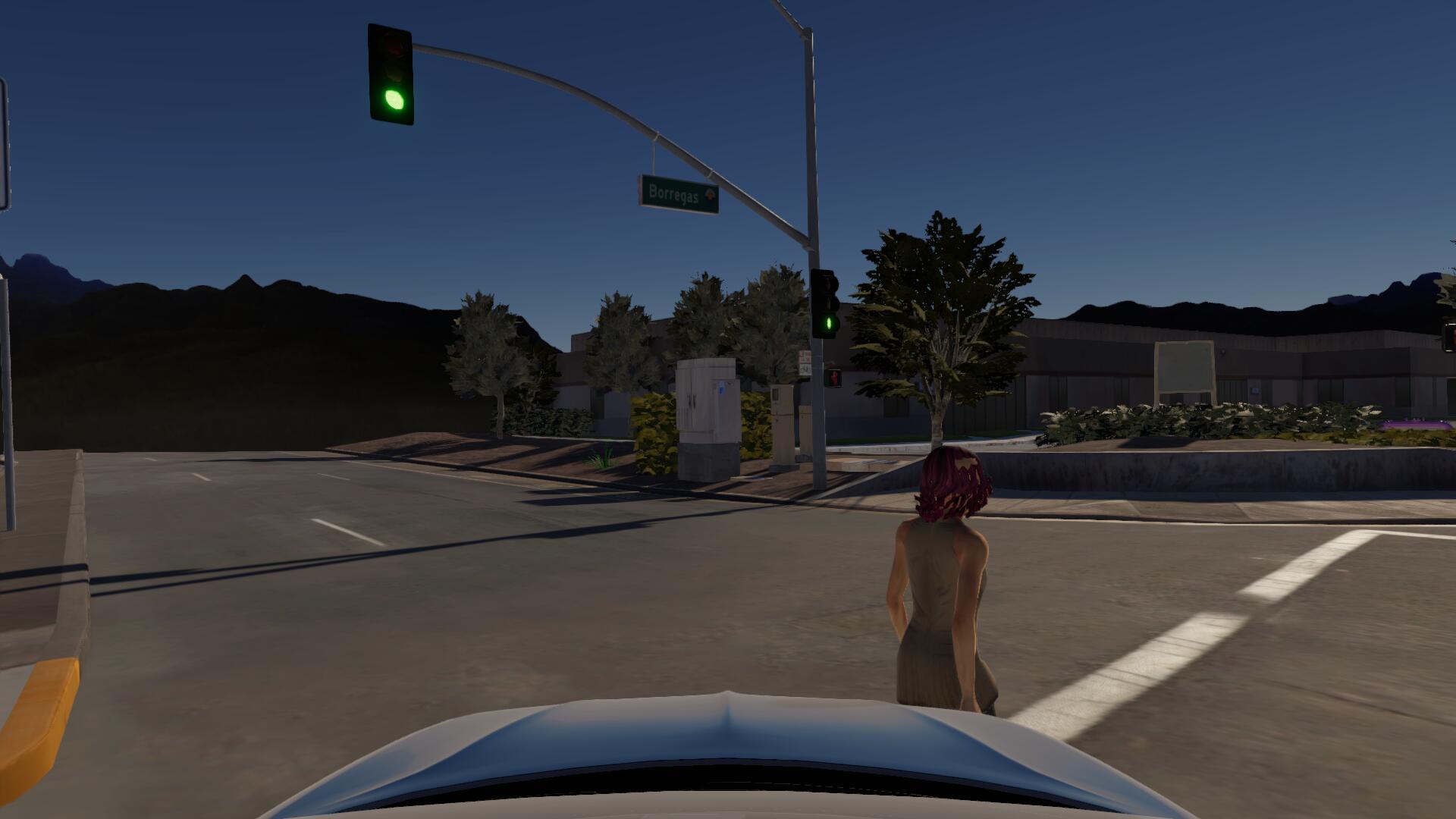}}
    
    \vspace{2mm}
    
    {\includegraphics[width=0.15\textwidth]{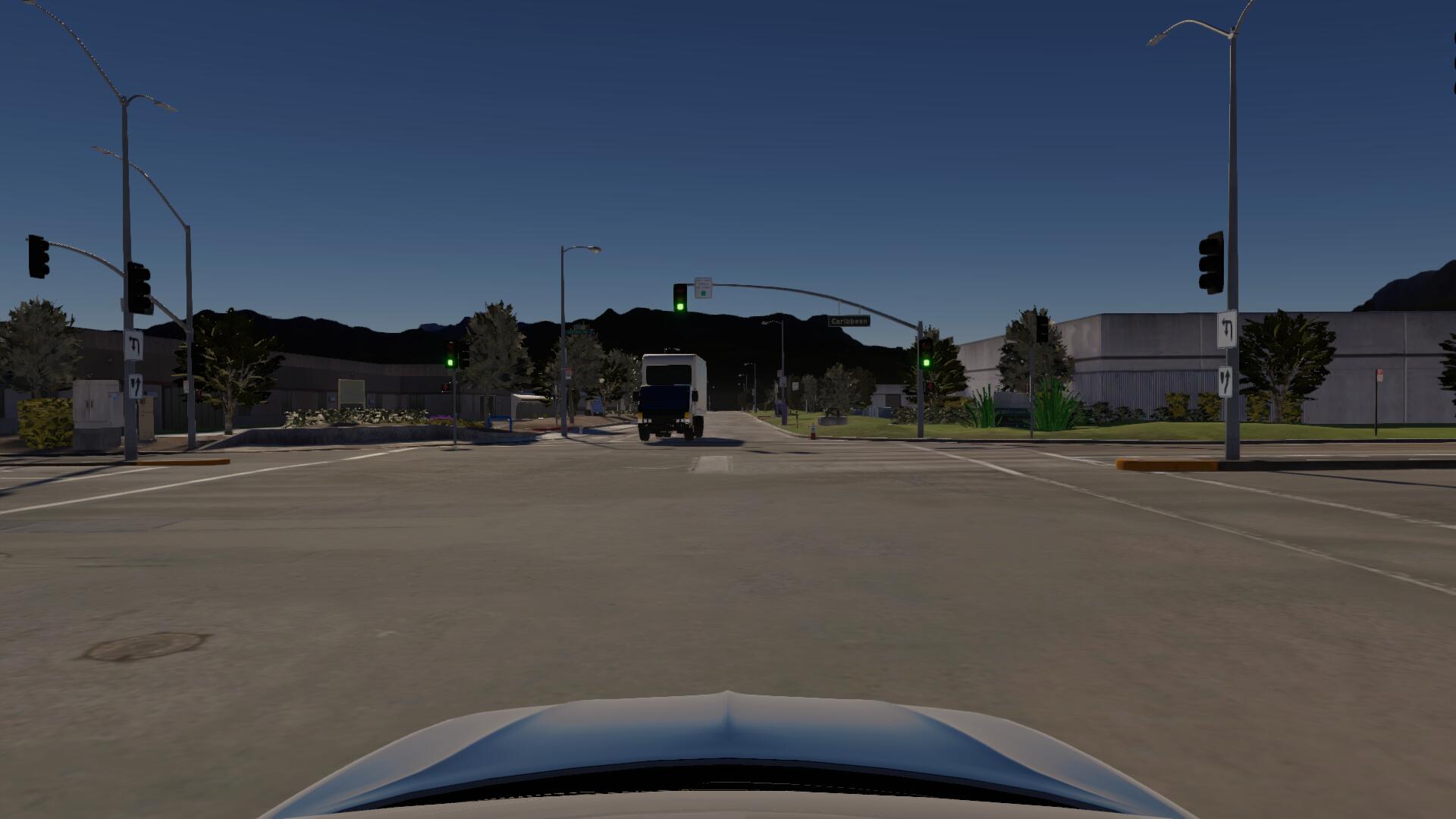}}
    {\includegraphics[width=0.15\textwidth]{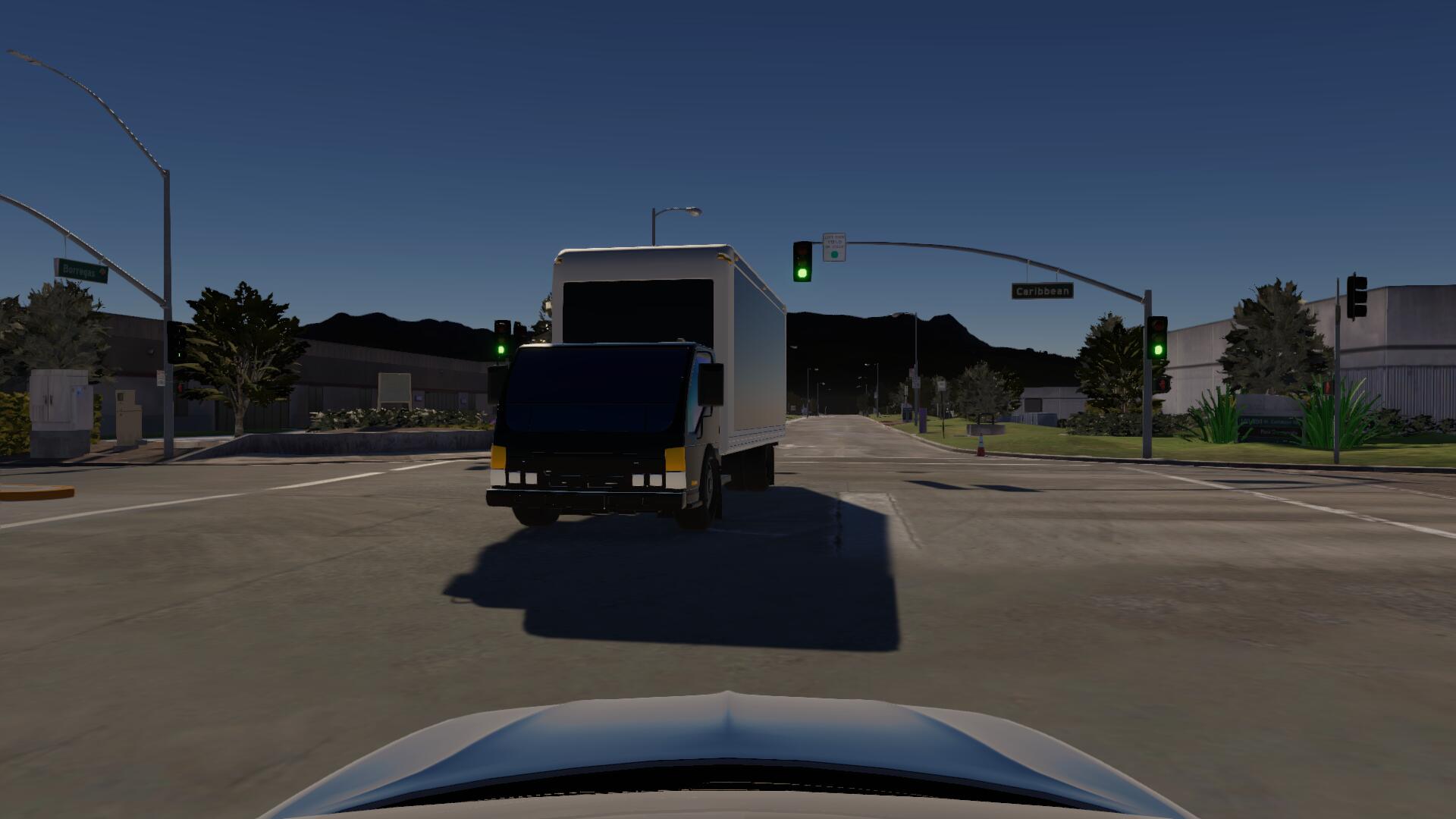}}
    {\includegraphics[width=0.15\textwidth]{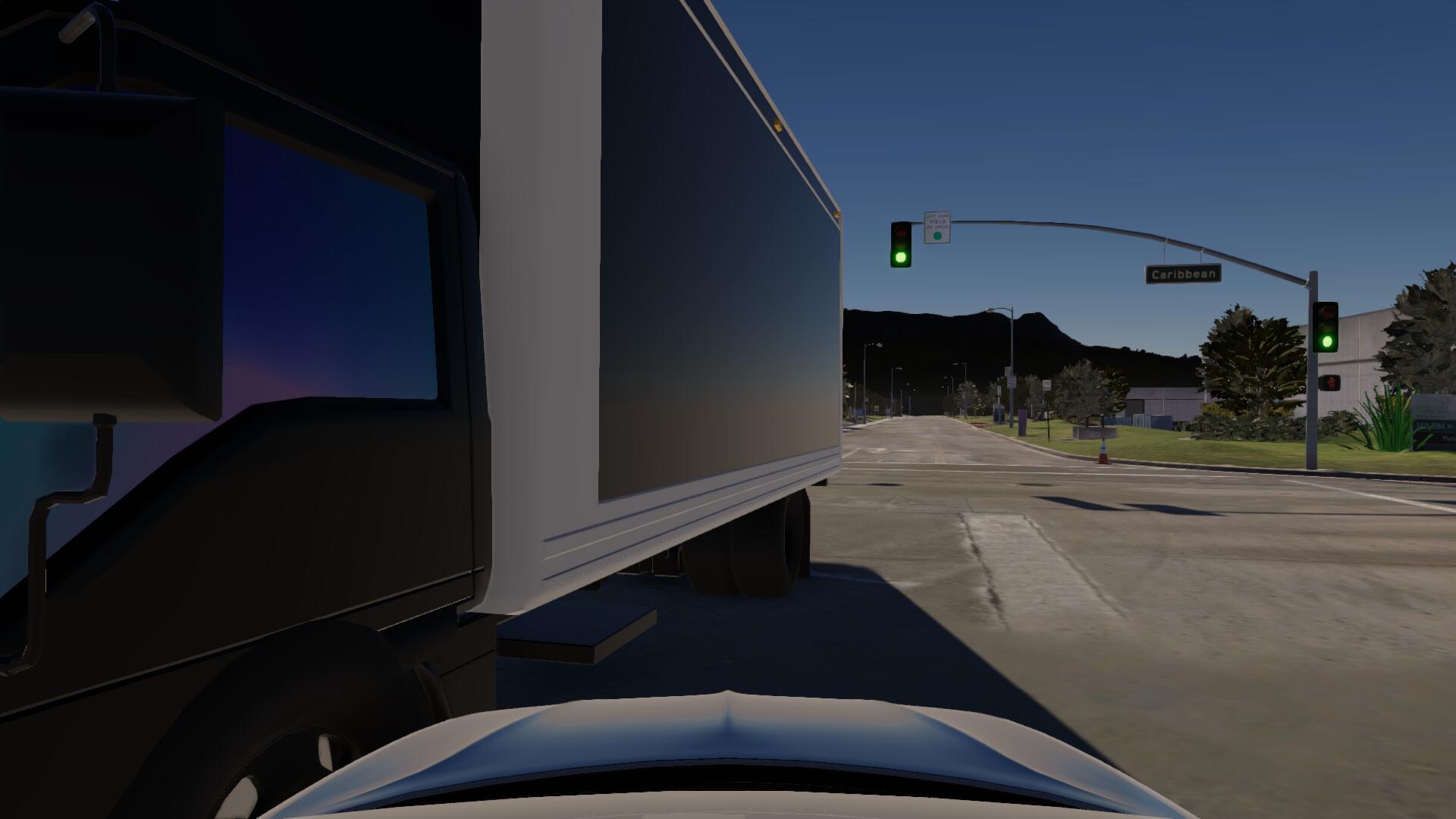}}
\caption{\textbf{\small{
\newedit{Two \bugs found for \apollo in \svl. (1st row) The ego-car turning left collides with a pedestrian crossing the street. (2nd row) The ego-car turning left collides with an incoming truck.}
}}}
\label{fig:more_bugs_demo_apollo}
\end{figure}

\remove{We have observed that \AVFUZZER finds much fewer \bugs because it has very limited diversity exploration. In particular, its default mutation rate is \sout{relatively }small and its local \GA starts with the mutated duplicates of the global best scenario vector so far, both of which limits diversity. If the global best scenario vector does not change after several generations, all the local \GA will start with its duplicates and thus further limit diversity. Moreover, its resampling process picks the scenarios farthest scenario vectors from existing ones but does not consider the distances among the selected scenario vectors, which results in restarting at a local cluster of scenario vectors with limited diversity.}
\Cref{fig:more_bugs_demo_apollo} shows two examplary Apollo \bugs found by \tool: the ego car turning left collides with a pedestrian crossing the street and an incoming truck, respectively. \new{They expose different functionality errors: fail to avoid a pedestrian and fail to avoid a truck, respectively. An investigation of the two violations identify their different causes. In the pedestrian collision case, the ego car's detection of the pedestrian is too late and thus the ego car does not have enough time to stop. In the truck collision case, the ego car detects the truck stably but does not plan its speed properly.}

\RS{4}{\newedit{\tool can generalize beyond \carla. In particular, it can find more unique \bugs than the baseline methods for \apollo in \svl.}}

%% file: body/8_related.tex
\section{Related Work}


\Cref{sec:background_test} presents the work most related to this paper. This section covers other peripheral works. 


\noindent
\textbf{Grammar-based Fuzzing.~}
Fuzzing produces input variations and tries to find failure cases for the software under test \cite{8449261, Dreossi2019VERIFAIAT}. Fuzzing tends to work well with relatively simple input formats such as image\new{ \cite{imagefuzz} or audio \cite{audiofuzz}.}\sout{, audio or video \cite{imagefuzz, audiofuzz, videofuzz, fuzzELF}.} For more complex input formats such as cloud service APIs \new{\cite{pythia20}}\sout{, XML parsers \cite{XMLfuzz}}\new{ or language compilers \cite{GCCfuzz}}, researchers often use grammar-based fuzzing \new{\cite{evgrammarfuzz, 9154602}} 
\sout{\cite{evgrammarfuzz, 10.1145/1375581.1375607, 10.5555/2362793.2362831, 9154602, 10.1145/1993316.1993532}}
to obey domain-specific constraints and narrow down the search space for producing effective and valid inputs. 

\noindent
\textbf{Language Specification and Testing.~}
OpenScenario~\cite{OpenScenario} is an open file format for describing the dynamic contents of driving simulations at a logical level \cite{ScenariosforDevelopment}, but it is at an early stage.  GeoScenario~\cite{GeoScenario} provides a language describing a \edit{\specifics} to be simulated; \edit{\cite{8722847} develops a simulation-based testing framework for \av{}. Neither provides a parametric search space that can be easily fuzzed.} In contrast, we parameterize \edit{\fss} 
that allows users to specify the range\new{ and distributions} of parameters\sout{, the}\new{ and their} constraints\sout{ between them, and their distributions} for automatically finding \bugs. 

%% file: body/9_discussion.tex
\section{Discussion \& Threats to Validity}



\begin{figure}[t]
\centering
    {\includegraphics[width=0.24\textwidth]{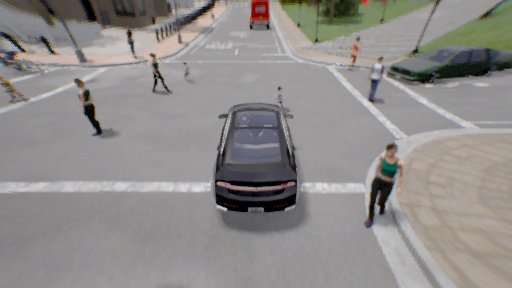}}
    \caption{\textbf{\small{An example of \bug in a high-dimensional scenario: the \av{} (controlled by lbc) collides with a child  crossing street.}}}
\label{fig:high_dim_bug_examples}
\end{figure} 

\noindent\new{\textbf{Realism.}} Our evaluation results are limited by the simulator implementations. Some reported \bugs might be due to interactions between the simulator and controller, \eg message passing delays, rather than the controller itself.  To mitigate this threat to internal validity, we experimented with two simulators (\carla and \svl) and four different controllers (lbc, pid1, pid2, \apollo). Further, to make the simulated crashes close to the real world, we construct \lss based on the most frequent pre-crash \fss from an NHTSA report.
The example shown in~\Cref{fig:high_dim_bug_examples} is a complex high-dimensional (328d) scenario with many agents. 
\new{Since to fully consider the temporal development (e.g., specifying the location of a vehicle at every time step), the search space can grow quickly and makes the searching process intractable, and during most accidents the movements of the involved vehicles/pedestrians can usually be decomposed into a couple of atomic behaviors, we currently consider one behavior development in \carla and only the initial state (e.g., location, orientation, and speed) in \svl. The integration of more temporal developments into \tool is relatively easy. Besides, given our fuzzing strategy's black-box nature, the additional behavior developments should only have limited influence.}

\noindent\new{\textbf{Road Infrastructures.} The road infrastructures considered in the current work are the default ones in the built-in maps in the simulators, which are mostly modeled based on the current road infrastructures in the United States. Different road infrastructures (e.g., those designed for deploying AVs) can influence the behavior of the AV under test \cite{Saeed2019, traffic_infra20}. However, the public road infrastructures with support for connected autonomous vehicles (CAVs) are not yet available and may not be available for quite some time. Nevertheless, there are not-connected AVs on conventional public roads now \cite{nhtsa_av_test_initiative_test_tracking_tool}, some of which led to fatal accidents \cite{tesladeaths}. Thus, it is necessary to study traffic violations by individual \av{}s on the current road infrastructures. We leave an exploration of road infrastructures with the support for CAVs for future work.
}

\noindent\new{\textbf{Unique Violations.}} 
The uniqueness of \bugs is hard to define precisely. We mitigate this threat to construct validity by extending the definition used in \cite{testing_vision18} with additional configurable parameters $th_1$ and $th_2$, enabling users to control uniqueness stringency. 
\newnew{A more desirable definition might be based on the internal system fault causing the violation. For example, two \bugs can be considered distinct if one is due to a failure of detecting a pedestrian for 2 seconds and the other is due to a sub-optimal tracking for 5 seconds). However, this is not feasible in our black-box testing setting where we assume no knowledge of the system under test. Besides, general methods to locate the root cause for a violation is itself an open question since it is non-trivial to assign the responsibility of a violation to different components of an AV at different time steps and the AV under test can have drastically different sub-components (e.g., lbc is an end-to-end neural network based system while \apollo is a modular based system).}
\newnew{Another}\remove{A more} desirable definition might be \newnew{search space }causal related, \eg only variables interacting with the ego car or that have an impact on ego car behavior count. However, efficiently determining the features contributing to a failure behavior is still an open challenge. One idea is to keep all other features fixed while changing the value of one feature and observe whether the failure behavior persists. If so, that feature can be potentially considered unrelated. This method faces some major limitations: First, as the number of features and the range for each feature become large, 
it is practically infeasible to conduct such analysis within a given time budget. Second, the features may not be independent and changing them one-by-one will 
miss the dependencies. Third, there is no consensus on quantifying if the causes of two failure cases are the same. For example, a car may collide with a pedestrian at slightly different locations for two simulations. Should we consider the cause to stay the same? It might be worth looking into the behavior of the controller’s internal states, which goes beyond the ability of a black-box testing framework.
Because of these challenges, we leave an in-depth study of this topic for future work.

\noindent\new{\textbf{Policy Implication:} 
Public officials should be educated on the severe implications of studies like our own, and urged to make a comprehensive \av safety testing standard. \av{}s must be shown to satisfy the safety requirements in the necessary testing process (e.g., having acceptably few traffic violations) in order to be allowed to deploy on the public roads.
}

%% file: body/10_conclusion.tex
\section{Conclusion}

We present \tool, a grammar-based fuzzing technique for finding \bugs in \av controllers during simulation-based testing. A \bug indicates a flaw in the controller that needs to be fixed. \tool leverages the simulator's API specification to generate inputs (seed scenes) from which the simulator will generate semantically and temporally valid \edit{\specificss}. It performs an NN-guided evolutionary search over the API grammar, seeking seeds that lead to distinct \bugs. Evaluation of our prototype implementation on \new{four}\sout{three} \av controllers shows that \tool successfully finds hundreds of realistic unique \bugs resembling complex real-world crashes and other driving offenses, outperforming the baseline methods. Furthermore, we \sout{capitalize on}\new{leverage} \bugs found\sout{ by \tool} to improve a learning-based controller's behavior on similar cases. \sout{Finally, we apply \tool on \apollo running in \svl to show its generalizability.} 



%% file: body/11_ack.tex
\section*{acknowledgement}
The work is supported in part by NSF CCF-1845893, NSF CCF-2107405, NSF IIS-2221943, NSF CCF-1815494, and DARPA/NIWC-Pacific N66001-21-C-4018. We want to thank Suman Jana and Dongdong She from Columbia University for valuable discussions.

%% file: body/appendix.tex
\appendices
\section{Details of \tool GA-based fuzzing}
Algorithm \ref{alg:ga-un-dnn-adv} shows the detailed implementation of our proposed algorithm \GAUNNNGRAD. The Neural Network part is highlighted in \textcolor{blue}{blue} and the gradient-based mutation is highlighted in \textcolor{red}{red}.

\begin{algorithm}
    \footnotesize
    \SetKwInOut{Input}{Input}
    \SetKwInOut{Output}{Output}
    
    \Input{\textbf{sampling()}, \textbf{evaluate()}, \textbf{max-gen}, \textbf{pop-size}, \textbf{generation-to-use-NN}, \textbf{candidate-multiplier}, \textbf{max-mating-iter}, \textbf{gradient-mutation-parameters}}
    \Output{unique-bugs}
    
    initial-population = sampling(pop-size)\;
    initial-unique-bugs, initial-objectives = evaluate(initial-population, [])\;

    generations = 1\;
    all-population = initial-population\;
    all-objectives = initial-objectives\;
    unique-bugs = initial-unique-bugs\;
    
    current-population, current-objectives = initial-population, initial-objectives\;
    
    \While {generations < max-gen}{
        mating-iter = 0;
        candidate-offspring = []\;
        remaining-size = pop-size $\times$ candidate-multiplier\;
        \While {mating-iter < max-mating-iter and remaining-num > 0}{
            mating-iter += 1\;
            parents = selection(current-population, remaining-num)\;
            new-candidate-offspring = crossover(parents)\;
            new-candidate-offspring = mutation(new-candidate-offspring)\;
            new-candidate-offspring = filtering(new-candidate-offspring, candidate-offspring, unique-bugs)\;
            
            candidate-offspring = merge(candidate-offspring, new-candidate-offspring)\;
            remaining-num = pop-size - len(candidate-offspring)\;
        }
        \If {remaining-num > 0}
        {
            remaining-candidate-offspring = sampling(remaining-num)\;
            candidate-offspring = merge(candidate-offspring, remaining-candidate-offspring)\;
        }

        \If {generations > generation-to-use-NN}
        {
            \textcolor{blue}{f = train-NN(all-population, all-objectives)}\;
            \textcolor{blue}{sorted-candidate-offspring = rank-by-confidence(candidate-offspring, f)}\;
            \textcolor{blue}{offspring = select(sorted-candidate-offspring)}\;
            \textcolor{red}{offspring = constrained-gradient-guided-mutation(offspring, f, gradient-mutation-parameters)}\;
        }
        
        new-unique-bugs, new-objectives = evaluate(offspring, unique-bugs)\;
        
        all-population = merge(all-population, offspring)\;
        all-objectives = merge(all-objectives, new-objectives)\;
        unique-bugs = merge(unique-bugs, new-unique-bugs)\;
        
        combined-population = merge(current-population, offspring)\;
        combined-objectives = merge(current-objectives, new-objectives)\;
        
        current-population, current-objectives = survival(combined-population, combined-objectives, pop-size)\;
        
        generations += 1\;
    }

    \Return unique-bugs

    \caption{\textbf{\small{\tool GA-based fuzzing (\GAUNNNGRAD) }}}
    \label{alg:ga-un-dnn-adv}
\end{algorithm}

\label{appendix}

\section{Additional \bugs examples}
\begin{figure}[h]
\centering
    {\includegraphics[width=0.15\textwidth]{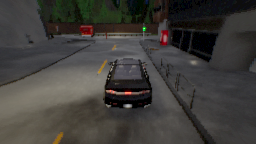}}
    {\includegraphics[width=0.15\textwidth]{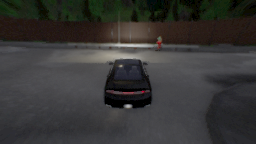}}
    {\includegraphics[width=0.15\textwidth]{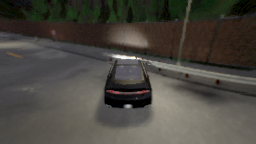}}
    
    \vspace{2mm}
    {\includegraphics[width=0.15\textwidth]{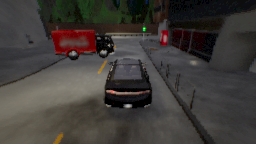}}
    {\includegraphics[width=0.15\textwidth]{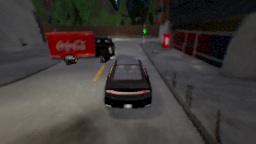}}
    {\includegraphics[width=0.15\textwidth]{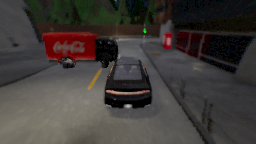}}
    
    \vspace{2mm}
    {\includegraphics[width=0.15\textwidth]{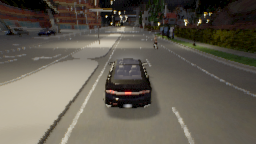}}
    {\includegraphics[width=0.15\textwidth]{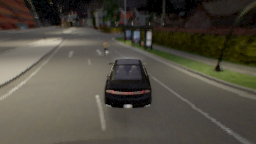}}
    {\includegraphics[width=0.15\textwidth]{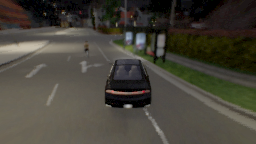}}

    {\includegraphics[width=0.15\textwidth]{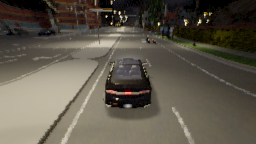}}
    {\includegraphics[width=0.15\textwidth]{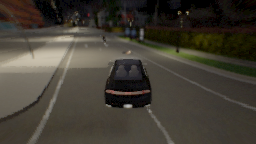}}
    {\includegraphics[width=0.15\textwidth]{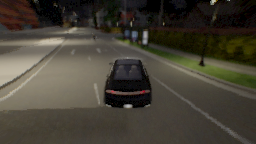}}
    
    \vspace{2mm}
    {\includegraphics[width=0.15\textwidth]{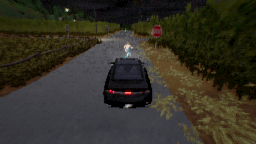}}
    {\includegraphics[width=0.15\textwidth]{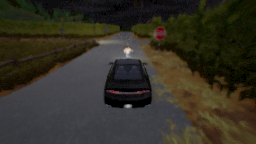}}
    {\includegraphics[width=0.15\textwidth]{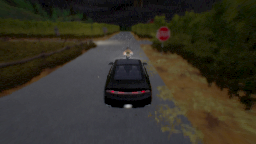}}
    
    \vspace{2mm}
    {\includegraphics[width=0.15\textwidth]{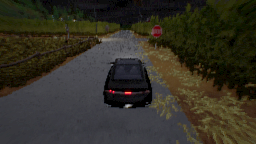}}
    {\includegraphics[width=0.15\textwidth]{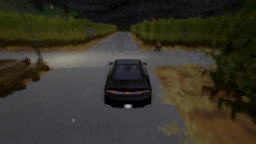}}
    {\includegraphics[width=0.15\textwidth]{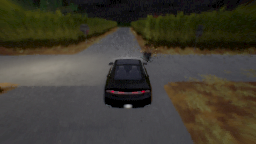}}
\caption{\textbf{\small{ \edit{Additional \bugs found in other scenarios:} (row1) lbc controller hits fencing \edit{in town01}. (row2) lbc controller hits a truck \edit{in town01}. (row3) lbc controller goes to the wrong lane \edit{in town03}. (row4) lbc controller hits a pedestrian lying on the road \edit{in town03}. (row5) lbc controller hits a pedestrian in front of it \edit{in town07}. (row5) lbc controller hits a cyclist crossing the street \edit{in town07}.}}}
\label{fig:more_bugs_demo}
\end{figure}

\Cref{fig:more_bugs_demo} shows more examples of found \bugs in other \ls. 

\section{Additional Figures}
\begin{figure}[h]
\centering
    {\includegraphics[width=0.4\textwidth]{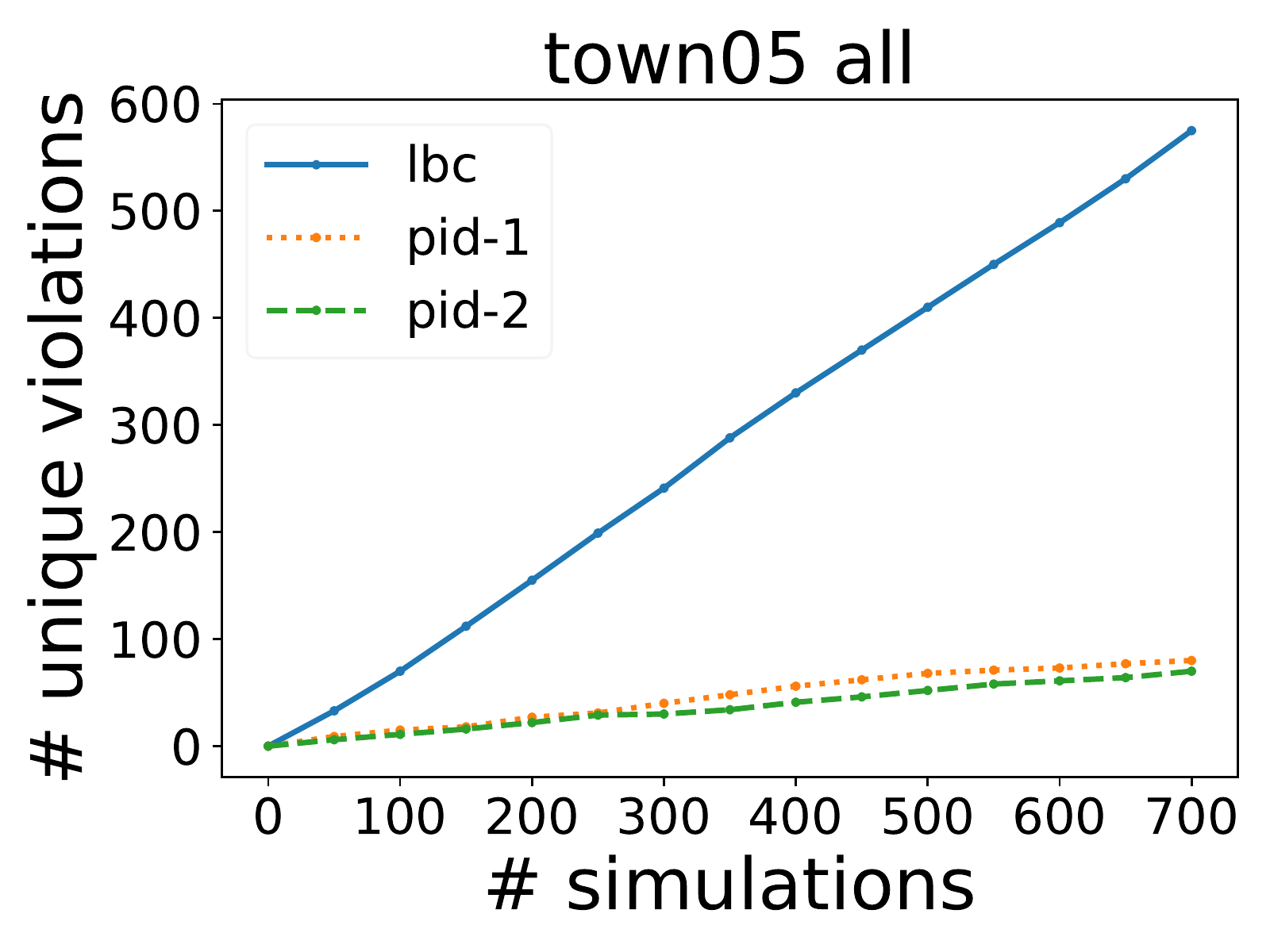}}
\caption{\small{\textbf{RQ1.~Unique \bugs (collision and out-of-road) found with three different controllers in a seed scenario where the ego car is approaching a junction while leading car slows down / stops. 
}}}
\label{fig:controllers}
\end{figure}

Figure \ref{fig:controllers} shows the number of all unique \bugs (including collision \bugs and out-of-road \bugs) found by \tool with \GAUNNNGRAD in town05 scenario over 700 simulations for each controller as described in RQ1.

\section{Details on Parameterizing the Fuzzable Fields}



\textbf{Constraining the API Grammar.} While fuzzing, each field can explore a large number of possible values, which significantly increases the search space and potentially leads to unrealistic scenarios. To generate more realistic scenario we enforce following two constraints on the API grammars.

\textit{(i)~Search Range.} While fuzzing, each field can explore a large number of possible values, especially for continuous fields, which significantly increases the search space and potentially leads to unrealistic scenarios. We restrict the search space by specifying a minimum and a maximum value of each fuzzable field (represented as $[min,max]$ in Listing~\ref{lst:ped_grammar}). A user can also optionally specify a clipped distribution for the field by appending the distribution's information: $[min, max, (distribution, mean, variance)]$, where the distribution is bounded by $[min,max]$. For example, in Listing~\ref{lst:ped_grammar}, a user mutates the pedestrian's location by sampling from a normal distribution with mean -103 and variance 10, and bounded by $-123$ to $-83$ range. With such parameterized field values, \tool uses \carla's built-in search function to look for defined nearby waypoints in the map and update the field values accordingly.  

\textit{(ii)~Constraints.} 
\tool's input space can be further constrained by providing additional conditions. For example, the distance between the ego-car and a car in front will be larger than a certain distance (see Listing~\ref{lst:ped_grammar}). 
A user can provide a list of constraints. 
For each constraint, three fields need to be specified. The first field `coefficients' is a list of coefficients to be multiplied before each of the selected field. The field `labels' is a list of corresponding fields where constraints are applied.  Finally, a `value' specifies the right-hand side of the inequality. The constraint is always encoded as $\leq$. 
Thus,  $coefficient[1] \times label[1] - coefficient[2] \times label[2] \leq value$. 

\section{Details and Illustration of objectives for Out-of-road Violatons}
\label{sec:illustration_of_wronglane_objective}
\begin{figure}[ht]
\centering
\includegraphics[width=0.2\textwidth]{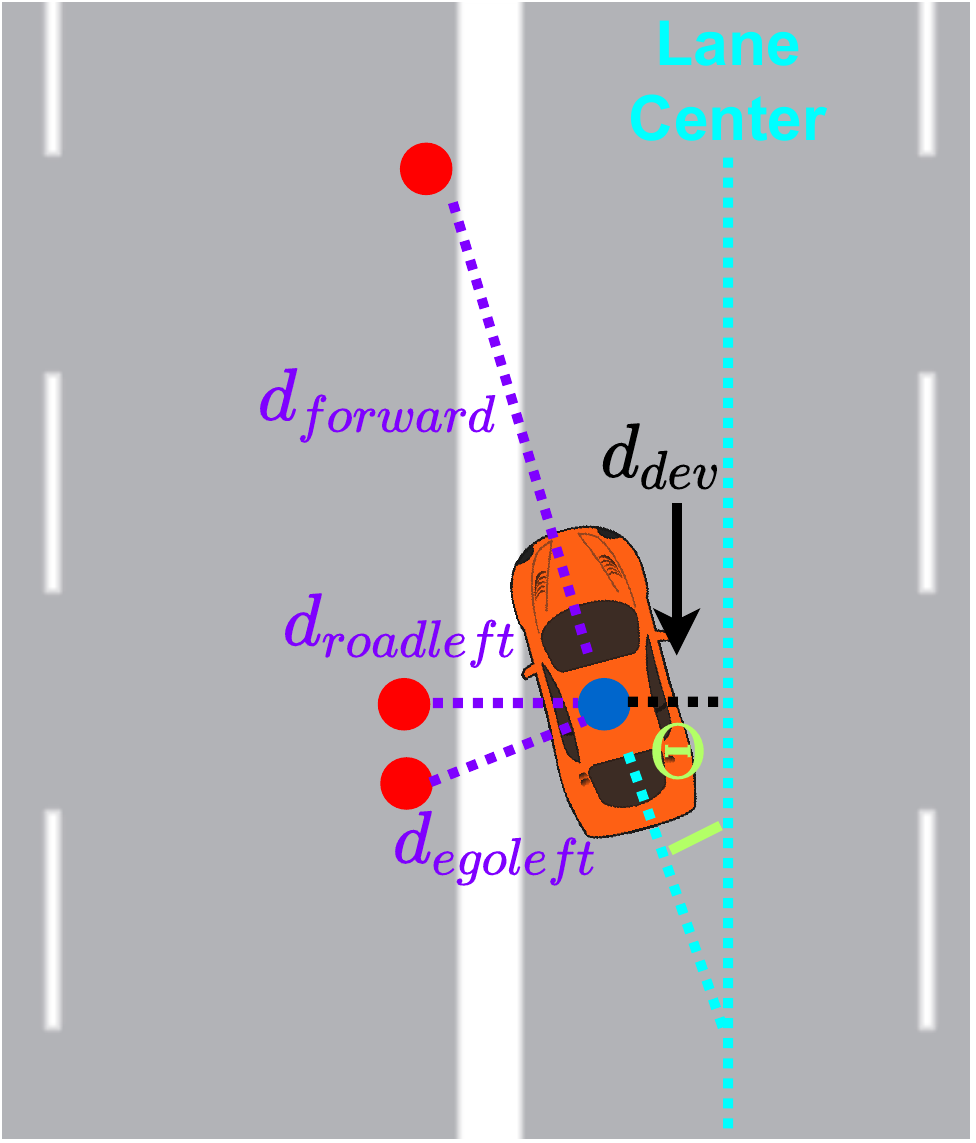}
\caption{\textbf{\small{Parameters computing the wrong lane objective function}}}
\label{fig:wronglane_objective}
\end{figure}

\Cref{fig:wronglane_objective} shows a car drives on a road. The angle between the ego-car's forward direction and the lane center is denoted $\Theta$ and the ego-car's center's distance from the lane center is denoted $d_{dev}$. $F_{deviation}$ is defined to be the product of the two i.e. $F_{deviation}=\Theta \times d_{dev}$.
$F_{wronglane}$ and $F_{offroad}$ measure the ego-car's distance from the closest lane of opposite direction or non-drivable region. In particular, $F_{wronglane}$ is the minimum of $d_{forward}$, $d_{roadleft}$, $d_{egoleft}$, $d_{roadright}$ and $d_{egoright}$ (the last two are not shown in \Cref{fig:wronglane_objective}), which denote the ego-car's distance along different directions \wrt the closest point that is on a road with different direction. $F_{offroad}$ is defined similarly except the closest point will be off road.


\section{Detailed results on percentage of \bugs being unique}
\label{sec:perc_unique_and_stat_test}

{

\begin{table}[ht]
\centering
\scriptsize
    \caption{\textbf{\small{Percentage of found \bugs that are unique}}}
    \label{tab:unique_percentage}
    \begin{tabular}{l|l|l|l|l|l}
    \toprule
        Map ID & GA-UN- & NSGA2 & NSGA2 & NSGA2 & AV\\
                  & NN-GRAD & -DT & -SM & -UN-SM-A & FUZZER\\
    \toprule
        town05 & $97.4\%$ & $56.8\%$	& $21.3\%$ & $96.4\%$ & $18.1\%$\\
        town01 & $100\%$ & $41.2\%$ & $23.1\%$ & $100\%$ & $9.8\%$\\
        town07 & $100\%$ & $67.5\%$  &  $66.0\%$ & $100\%$ & $16.2\%$\\
        town03 & $99.9\%\%$ & $79.4\%$ & $54.0\%$ & $100\%$ & $24.0\%$\\
    \bottomrule
    \end{tabular}

\end{table}
}




{

}

\Cref{tab:unique_percentage} shows the detailed results of the percentage of found \bugs by each method that are unique. Without surprise, \GAUNNNGRAD and \NSGAUNSMA have much higher percentages than the other two.

\section{More details on \apollo in \svl}
\svl officially only provides \apollo (modular testing) in which the perception, camera and traffic light modules of \apollo are not activated. Instead, the ground-truth information of other objects as well as traffic light are provided. To test the perception and camera modules as well as other modules, we created a version which has these modules activated. However, the traffic light modules cannot be successfully activated and it is a known issue on \svl github repo. As a workaround, the ground-truth traffic light information are fed into \apollo. In this way, we can test all modules of \apollo except the traffic light module.

\section{A Logical Scenario Example}
\label{sec:search_space_details}

\newnew{In this section, we show the details of the \ls "Turning right while leading car slows down/stops" we used in the current work. \Cref{tab:search_fields} shows its detailed search space, which has $26$ dimensions in total. Weather and lighting consists of a series of combination of weather and lighting condition (e.g., clear noon, rainy sunset, cloudy night, etc.). The detailed specification can be found on \carla Python API \cite{carla}. The number of NPC pedestrians, NPC vehicles, and static objects can also be searched. However, for this particular \ls, they are fixed such that the leading vehicle and a pedestrian always exist and no static object can influence the interaction between the ego car and the NPC vehicle/pedestrian.}

\newnew{Each NPC pedestrian's type (e.g., adult, child, etc.), initial location (i.e., initial x and initial y) and orientation (i.e., initial yaw) are also considered. Trigger distance controls the distance between the pedestrian and the ego car such that when the car is closer to the pedestrian than the trigger distance the pedestrian starts to move at target speed for the specified distance to travel.}

\newnew{Similar to a NPC pedestrian, each NPC vehicle also has fields like type, initial location and orientation, trigger distance, target speed, and distance to travel. It additionally has a color field controlling its color. Besides, if its field of waypoint follower is set to 1, the NPC vehicle will plan and move to the target location (with target x, target y, target yaw) rather than just moving along the initial orientation for distance to travel. It also has the field of avoid collision. When avoid collision is set to 1, the NPC vehicle will stop if a collision is likely to happen given its current velocity and other vehicle's current velocity.}

\newnew{The initial location and end location of the ego car also need to be specified. They are not explicitly shown in \Cref{tab:search_fields} since they are not searched. For the current \ls, they are hard-coded. Besides, all the location related fields are relative coordinates with respect to some center coordinates. In this \ls, the middle point along the route between the initial location and the end of location of the ego car is used for the center of both the pedestrian and the vehicle.}

\newnew{The details for other \lss can be found in our publicized source code \cite{ziyuan_zhong_2022_6399383}.}

\begin{table}[ht]
\scriptsize
    \centering
    
    \begin{tabular}{l|l|l|l}
    \toprule
        object & property & data type & range\\
    \toprule
    background & road fraction & continuous & [0.7, 0.9] \\
    & weather and lighting type & discrete & [0, 14] \\
    
    \midrule
    agent general & number of pedestrians & discrete & [1, 1] \\
    & number of vehicles & discrete & [1, 1] \\
    & number of static & discrete & [0, 0] \\
    
    \midrule
    $\textrm{pedestrian}_i$ & type & discrete & [0, 3] \\
    $i\in$ & initial x & continuous &[-10, 10] \\
    $\{1,...,$ & initial y & continuous & [-10, 10] \\
    $\textrm{number of pedestrians}\}$ & initial yaw & continuous & [0, 360] \\
    & trigger distance & continuous & [10, 50] \\
    & target speed & continuous & [0, 4] \\
    & distance to travel & continuous & [0, 50] \\
    
    \midrule
    $\textrm{vehicle}_i$ & type & discrete & [0, 10] \\
    $i\in$ & color & discrete & [0, 4] \\
    $\{1,...,$ & initial x & continuous & [-0.5, 0.5] \\
    $\textrm{number of vehicles}\}$ & initial y & continuous & [-12, -5] \\
    & initial yaw & continuous & [0, 360] \\
    & initial speed & continuous & [2, 5] \\ 
    & trigger distance & continuous & [5, 12] \\
    & target speed & continuous & [0, 2] \\
    & if waypoint follower & discrete & [0, 1] \\
    & distance to travel & continuous & [5, 30] \\
    & avoid collision & discrete & [0, 1] \\
    & target x & continuous & [-20, 20] \\
    & target y & continuous & [-20, 20] \\
    & target yaw & continuous & [0, 360] \\
    \bottomrule
    \end{tabular}
    \caption{\small{Details of the search space.} \label{tab:search_fields}}
\end{table}